## JGR Solid Earth RESEARCH ARTICLE 10.1029/2021JB023071

# Present-Day Surface Deformation of Sicily Derived From Sentinel-1 InSAR Time-Series

**Maxime Henriquet[1,2], Michel Peyret[1], Stéphane Dominguez[1], Giovanni Barreca[3], Carmelo Monaco[3], and Stéphane Mazzotti[1]**

[1]Géosciences Montpellier, Université de Montpellier, Montpellier, France, [2]Aix-Marseille Université, CEREGE, Aix-en- Provence, France, [3]Dipartimento di Scienze Biologiche, Geologiche e Ambientali, Sezione di Scienze della Terra, Università di Catania, Catania, Italy

**Key Points:**

- Processing Sentinel-1 data from 2015 to 2020 provide the first large-scale surface velocity field of Sicily at a high spatial resolution.
- The estimated low deformation rates in Sicily result from the complex interplay between permanent and transient processes.
- The most striking tectonic features concerns the Cefalù-Etna seismic zone, Mount Etna dynamics and the eastern coast of the Hyblean Plateau.

**Correspondence to:** M. Henriquet, henriquet@cerege.fr

**Abstract:** The Quaternary geodynamics of the Central Mediterranean region is controlled by the migration of narrow orogenic belts within the slow Nubia-Eurasia plate convergence. As testified by the occurrence of major volcanic and seismic events, the Eastern Sicilian Margin is presently one of the most active regions. Using a Permanent-Scatterer approach, we process Sentinel-1 satellite images acquired from 2015 to 2020 to provide an island-wide quantification of surface displacements at a high spatiotemporal resolution. We then convert the calculated mean surface velocities along the ascending and descending satellite line of sight into the ITRF2014 reference frame by using GNSS velocity data derived from regional stations. The resulting pseudo-3D velocity field mainly highlights a general uplift of about 1.5 ± 0.5 mm/yr of the Nebrodi-Peloritani range and its differential motion with respect to mainland Sicily along the Cefalù-Etna seismic zone. Permanent/Persistent-Scatterer (PS) vertical velocities in the Eastern Hyblean region reveal a long wavelength eastward down-bending of the margin, including the inferred epicentral area of the 1693 Noto earthquake. Compared to Quaternary coastal uplift rates, these results confirm the relative low activity of Western Sicily, a potential slow uplift of South-Central Sicily and a significant discrepancy along the Eastern Hyblean margin were PS-derived vertical velocities that appear 2–3 mm/yr lower than the Quaternary rates. Over the 2015–2020 timespan, transient processes are also captured, notably on Mount Etna, showing both magmatic pressurization uplift and collapse of the eastern flank, but also all over Sicily where numerous gravitational mass movements and anthropogenic ground subsidence are detected.

**Plain Language Summary:** The recent geodynamics of Central Mediterranean is controlled by the migration of narrow orogenic belts within the slow Nubia-Eurasia plate convergence. As testified by major volcanic and seismic events, the Eastern Sicilian Margin is a region of strong tectonic activity. Using a Permanent-Scatterer approach, we process Sentinel-1 data to provide an island-wide quantification of surface displacements from 2015 to 2020. We convert the calculated mean surface velocities along the satellite line of sight into the ITRF2014 reference frame using GNSS data. The resulting velocity fields highlight a general uplift of the Nebrodi-Peloritani range and its differential motion relative to mainland Sicily along the Cefalù-Etna seismic zone. Permanent/Persistent-Scatterer (PS) vertical velocities in the Eastern Hyblean region reveal a long wavelength eastward down-bending of the margin including the inferred epicentral area of the 1693 Noto earthquake. These results are compared to Quaternary coastal uplift rates, confirming the relative low activity of Western Sicily, a potential slow uplift of South-Central Sicily, and a significant discrepancy along the Eastern Hyblean margin where PS vertical velocities are lower than the Quaternary rates. Transient processes are also captured, notably on the Etna volcano, but also all over Sicily where numerous landslides or anthropogenic ground subsidence are captured.

## 1. Introduction

Since the middle of the Cenozoic, the Mediterranean geodynamics is dominated by the growth of narrow orogenic systems driven by fast slab retreat, interacting with the slowly converging Nubian and Eurasian plates (e.g., Barreca et al., 2010; Doglioni, 1991; Doglioni et al., 2007; Malinverno & Ryan, 1986; Royden & Faccenna, 2018). The Sicilian orogen (Figure 1a) is the result of the Late Miocene and younger collision between the Calabro-Peloritani block and the African continental passive margin (Henriquet et al., 2020 and references therein), while the Calabrian Arc follows the southeastward retreat of the Ionian oceanic plate (Faccenna et al., 2005; Goes et al., 2004).

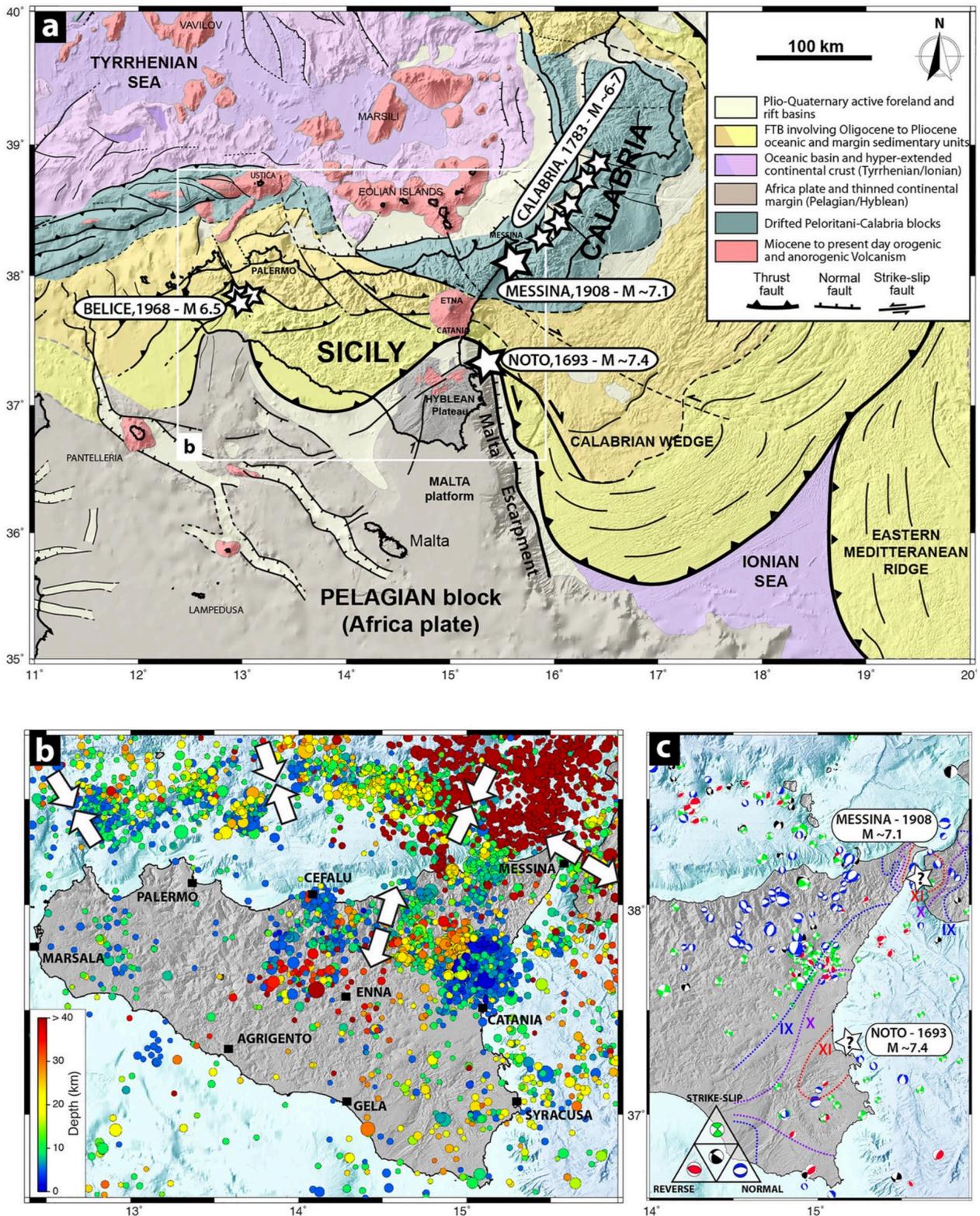

**Figure 1.** Geological and geodynamic context of Central Mediterranean and seismicity of Sicily. (a) Simplified neotectonic and structural map of Sicily (modified from Henriquet et al., 2020). Main Plio-Quaternary faults are outlined in black, and yellow stars show the approximate locations of major historical and instrumental earthquakes. Instrumental seismicity shows hypocentral locations of $M > 2.5$ events from 1985 to 2021 (b, from http://cnt.rm.ingv.it) and focal solutions of $M > 3$ events from 1999 to 2020 in Eastern Sicily, excepted earthquakes related to the Ionian subduction (c, from http://ct.sismowebingv. it). White arrows represent the general tectonic context from D'Agostino et al. (2008). Macroseismic intensity contours of the Messina 1908 and the Noto 1693 earthquakes are from Ridente et al. (2014) and Barbano (1985).

Four main tectono-stratigraphic domains are attributed to the Sicilian orogen (Figure 1, e.g., Henriquet et al., 2020): (a) To the northeast (Peloritan and Calabrian mountain ranges), the Calabro-Peloritani block corresponds to upper-plate continental remnants derived from the European margin. (b) In Central and Northern Sicily, the Alpine Tethys accretionary wedge relics come from the offscraping of the neo-Tethyan sedimentary cover during the Oligocene-Early Miocene subduction phase. (c) All over Sicily (excepted the southeastern and northeastern regions), the core of the fold-and-thrust belt is made of an imbricated Meso-Cenozoic platform and deep-water units detached from the continental African passive margin during the continental subduction and collision phases that took place since the Middle Miocene. (d) To the southeast, the Hyblean domain corresponds to the foreland against which the Sicilian Chain collided.

The present-day kinematics of Sicily has been investigated using geodetic measurements (e.g., Angelica et al., 2013; Bonforte & Guglielmino, 2008; Bonforte et al., 2016; D'Agostino et al., 2011; Devoti et al., 2008, 2011, 2014, 2017; Mastrolembo Ventura et al., 2014; Mattia et al., 2008, 2009, 2012; Serpelloni et al., 2010, 2013; Palano et al., 2012) and stress orientations derived from earthquake focal mechanisms (e.g., Montone et al., 2012; Pondrelli et al., 2006; Soumaya et al., 2015). Over the last two decades, the expansion of GNSS networks has resulted in numerous compilations of geodetic velocity fields covering the whole Italian peninsula (e.g., Devoti et al., 2008, 2011, 2017; Angelica et al., 2013; Mastrolembo Ventura et al., 2014; Serpelloni et al., 2013).

These data sets reveal (a) the independent motion of the Sicilian foreland (Hyblean Plateau) relative to the Nubian plate (mainland Africa), characterized by a low extensional rate of 1–2 mm/yr accommodated in the Sicily Channel by NW-SE-oriented structural depressions, such as Pantelleria, Linosa, and Malta grabens (e.g., Belguith et al., 2013; Palano et al., 2012; Soumaya et al., 2015); (b) ~3 mm/yr of extension in northeastern Sicily within the Nebrodi-Peloritani range and across the Messina strait (e.g., Cultrera et al., 2017; D'Agostino & Selvaggi, 2004; Mastrolembo Ventura et al., 2014); (c) the contraction between the Hyblean foreland and the south of the Nebrodi Mounts (e.g., Bonforte et al., 2015; Mastrolembo Ventura et al., 2014; Mattia et al., 2012; Musumeci et al., 2014) with a shortening rate of 2–3 mm/yr (Mastrolembo Ventura et al., 2014); and (d) the compression in the southern Tyrrhenian Sea as evidenced by the ~2 mm/yr of N-S shortening between Ustica and the northern Sicilian coast (e.g., Mastrolembo Ventura et al., 2014; Palano et al., 2012; Soumaya et al., 2015).

Despite relatively low current deformation rates (<3 mm/year, compared to the 5 mm/yr African-Eurasian convergence motion, e.g., D'Agostino et al., 2011), the activity of the Sicilian region remains in general agreement with the known Plio-Quaternary geodynamic setting (Figure 1a). Locally, post-Middle Pleistocene stress reorganization patterns are identified in the Hyblean foreland, with strike-slip reactivation of inherited normal faults (e.g., Cultrera et al., 2015); northeastern Sicily, with the activation of dextral fault zones separating north-eastern Sicily from mainland Sicily (e.g., Barreca et al., 2019; Billi et al., 2006; Cultrera et al., 2017; De Guidi et al., 2013; Pavano et al., 2015); and in Calabria, where uplift pulses imprint the landscape morphology of the Aspromonte Massif (Robustelli, 2019).

This complex kinematics, particularly along Eastern Sicily, has a significant impact on the seismic activity of this region (Figures 1a and 1b), which comprises the most destructive historical events recorded in Italy, such as the Noto earthquake in 1693 (macroseismic intensity of XI and Mw ~ 7.4) that struck Central and Eastern Sicily (Boschi, 2000). The Messina 1908 earthquake is the second major historical event that caused signifi- cant damages and the death of more than 80,000 people on both sides of the Messina Strait (Mw ~ 7.1; Aloisi et al., 2012; Baratta, 1910; Barreca et al., 2021; Mercalli, 1909). In 1783, a series of earthquakes of magnitude 5–7 also struck Eastern Calabria (Boschi, 2000; Jacques et al., 2001). Lastly, Western Sicily was affected by the Mw ~ 6 Belice earthquake in 1968 (e.g., Anderson & Jackson, 1987; Bottari, 1973; Monaco, Mazzoli, & Tortorici, 1996), which led to a revision of the seismic hazard in this region.

The correlation of the regional GNSS velocity field with known active faults remains challenging in Sicily (Mastrolembo Ventura et al., 2014; Palano et al., 2012). Away from areas affected by volcanic processes, present-day horizontal and vertical displacements are low (<2–3 mm/yr), often close to (or even below) the GNSS velocity resolution. To date, only a few large-scale studies have intended to estimate the vertical component of the surface velocity field (e.g., Anzidei et al., 2021; Devoti et al., 2017; Serpelloni et al., 2013). However, despite a good GNSS network density, notably in the eastern region, the active deformation zones remain poorly constrained. Moreover, the lack of clear fault surface ruptures related to the strongest historical earthquakes (Noto 1693, Messina 1908, and Belice 1968) makes the seismogenic source of these events still debated (e.g., Barreca et al., 2014; Cultrera et al., 2015; Gambino et al., 2021; Meschis et al., 2019; Visini et al., 2009), strongly limiting seismic hazard assessment.

The processing and analysis of InSAR time series offer key data sets to fill the gaps left by sparse GNSS measurements. Such a tool is commonly used to quantify coseismic surface ruptures (e.g., Xu et al., 2020), inter-seismic loading (e.g., Cavalié et al., 2008), or active tectono-magmatic processes (e.g., De Novellis et al., 2019; Monaco et al., 2021). Up to now, the use of InSAR data in Sicily has been limited to local investigations, mostly dedicated to the dynamics of Mount Etna (e.g., Bonforte et al., 2011; Borgia et al., 2000; De Novellis et al., 2019; Doin et al., 2011; Froger et al., 2001), and to investigate the kinematics of the Hyblean region (Anzidei et al., 2021; Canova et al., 2012; Vollrath et al., 2017), and south-western Sicily (Barreca et al., 2014). It was also recently used at a large spatiotemporal scale to quantify transient surface deformations from 1992 to 2014 over the whole Italian territory (Costantini et al., 2017; Piombino et al., 2021). However, newer high resolution spatiotemporal data from Sentinel-1 satellites can offer more precise mean velocity fields, suitable for tectonic interpretations in regions affected by low to moderate deformation rates.

The present study aims to provide and discuss the first geodetic velocity field covering the whole Sicily Island, derived from Sentinel-1 InSAR time series over the 2015–2020 period. Mean velocities along the satellite line of sight (LOS) in both ascending and descending passes are obtained using the Persistent-Scatterer approach (Hooper et al., 2004). The East and Up components of the velocity field are then reconstructed and adjusted to reliable GNSS velocities. The PS-InSAR analysis relies on the construction of time series for each selected PS, which allows the estimation of an average velocity by assuming a stable displacement in time (such as regional tectonic loading). In the case of transient processes, this average velocity has little meaning, but the time series documents the evolution of the deformation in detail. First-order kinematics of Sicily is then discussed at the local and regional scales to refine tectonic interpretations and seismic hazards on the Hyblean Plateau, the Etnean region, the Nebrodi-Peloritani range, and Western Sicily. Analyzing the InSAR time series or comparing the retrieved velocities with long-term rates highlights known transient processes, such as volcanic dynamics, anthropogenic subsidence, gravitational mass movements, and potentially inter-seismic elastic loading.

## 2. Ground Deformation Map From PS-InSAR Analysis of Sentinel-1 Data

### 2.1. Methodology Overview

InSAR time series analyses can be classified into two major technics, depending on the underlying definition of what is stable with time. One type of approach, working at a full spatial resolution with interferometric time series built from a single primary image, identifies pixels with stable backscattering properties in both amplitude and phase (e.g., Ferretti et al., 2001; Hooper et al., 2004). These stable pixels, called Permanent/Persistent-Scatterers (PS), generally correspond to human-made structures or rock outcrops. One single dominant scatterer within the pixel may suffice to select it as a PS, leading to a spatial resolution that can be smaller than the pixel size. An alternative to PS-InSAR methods is called SBAS (Small-Baseline, e.g., Bernardino et al., 2002). Its principle, rather than detecting individual dominant scatterers at a high spatial resolution, consists in increasing the pixel coherence via spatial averaging (so-called "multilooking") at the expense of the spatial resolution. Using this technic, a network of interferograms with small temporal and perpendicular baselines is built, and a mean velocity map is inverted. Finally, a combination of these two approaches can be performed, taking advantage of both: the full resolution of the PS-InSAR technic and the high data redundancy of the SBAS technic (Ferretti et al., 2011; Hooper, 2008).

In this study, the PS-InSAR version of the StaMPS processing chain has been implemented, since the primary objective was to map ground deformation at the highest spatial resolution. As shown hereafter, despite the high vegetation coverage of some regions in Sicily, the spatial coverage of the detected PS is excellent. Hence, no complementary SBAS approach has been done, even though it would be interesting to compare both approaches from a methodological point of view.

The LOS mean velocity fields derived from InSAR time series have an arbitrary zero reference. To overcome this issue, we first project reliable GNSS 3D velocity measurements along the satellite LOS. Then, we calculate the LOS velocity misfit between the InSAR-derived velocity and the GNSS-derived velocity at each site and fit a planar ramp through the velocity misfit field. This average misfit is then subtracted from the PS velocity field to adjust it to the GNSS reference. This correction contributes to attenuate the small residual orbital errors affecting the Sentinel-1 acquisitions and provides a common reference frame to compare the PS and GNSS velocity fields (Yang et al., 2019). While it is usually accepted that a linear ramp correction provides good results in the InSAR-GNSS adjustment (Pathier et al., 2003; Yang et al., 2019), we cannot rule out that minor additional uncertainties may be added to long-wavelength signals.

Due to the mono-dimensional characteristic of the radar measurement (LOS motion), the calculation of 3D ground displacements (or velocities) requires the combination of ascending and descending InSAR measurements with additional constraints and assumptions. One approach consists in using a spatially sparse set of geodetic measurements (typically GNSS). However, the incorporation of these complementary constraints requires some interpolation. This can be done explicitly (e.g., Catalão et al., 2011; Gudmundsson et al., 2002; Samsonov et al., 2008; Wang et al., 2012) or implicitly via the use of local mechanical behaviors (e.g., Guglielmino et al., 2011). In this study, the North component of the ground displacements is very small compared to the East and Up components when projected along the satellite LOS (heading angle ~10° with respect to the North-South direction). Considering the GNSS velocity field (Figure S1 in Supporting Information S1), we get a maximum of North-velocity gradient of about 5 mm/yr (Figure S2 in Supporting Information S1), which projects to about 0.5 mm/yr along the LOS (Figures S3–S6 in Supporting Information S1). Although this small value is below the mean PS velocity uncertainty, we estimate and subtract it to the PS mean LOS velocities to get a linear system with two unknowns (East and Up velocities) and two measurements (PS velocities in ascending and descending passes). The interpolation of the north-velocity component of the sparse GNSS network is based on a very smooth interpolator (continuous curvature splines with null tension, Figure S2 in Supporting Information S1) to avoid any unrealistic short-wavelength velocity biases. This interpolation can certainly be improved, but modifying the interpolation method will lead to East- and Up-component changes that will not exceed a fraction of a mm/yr (see the Figures S7 and S8 in Supporting Information S1 for a comparison with the assumption that the North component is constant). Hence, in the following, the East and Up components are extracted from the PS velocity field corrected from the North component of the GNSS velocity field.

As developed in many studies (e.g., Guglielmino et al., 2011; Ng et al., 2011; Spata et al., 2009; Wright et al., 2004), this assumption reduces the problem to a 2D estimation (e.g., East and Up components) from 2D measurements using ascending and descending geometries. In principle, the optimal decomposition frame for 2D velocity extraction from LOS velocity fields is the bisectors of the heading angles (349.70° and 190.00°), which deviate from the E-W and N-S axes by only 0.15°. For clarity and because the deviation from a standard reference frame is very small, we did not choose to decompose the LOS velocities on such bisectors and kept the common E-W and N-S horizontal frames. Practically, ascending and descending PS velocity fields, which were adjusted to the GNSS velocity field, are resampled along a common uniform geographical grid. The grid-sampling interval is defined to a value slightly larger than the SAR image resolution to capture ascending and descending PS information at one single location. We chose a sampling interval of 30 m, averaging PS velocities in a radius of 500 m. Then, at every pixel with both ascending and descending velocities ($v_{ac}$ and $v_{desc}$), we solve for the East and Up components ($v_e$ and $v_{up}$) in a weighted least-square sense by using the following equation:

$$\begin{pmatrix} lv_e \\ v_{up} \end{pmatrix} = -\left[ P'^t.S.P' \right]^{-1} P'^t.S \begin{pmatrix} lv_{asc} \\ v_{desc} \end{pmatrix}$$

where $P$ is a $2 \times 2$ matrix made of the coefficients of projection along the LOS, and $S^{-1}$ is the covariance matrix of the PS velocity measurements. Assuming that the PS velocities estimated in the two different geometries are independent, $S$ is diagonal.

The resolution matrix $T$ takes the following form:

$$T = P'^t.S.P'$$

Leading to the quantification of the uncertainty of the East and Up components as the diagonal elements of matrix T. The non-diagonal element of $T$ (which is a symmetric matrix), normalized by the square root of the product of the diagonal elements, provides the correlation between Up and East uncertainties.

Finally, in places where adjacent radar frames with similar satellite direction pass overlap, one may have two distinct measurements obtained with two distinct LOS. One LOS corresponds to the far range of the radar acquisition (incidence angle ~45°), while the other corresponds to the near range of the adjacent radar acquisition (incidence angle ~18°). In that case, Equation 1 can be extended using an additional measurement ($v_{ac}$ and $v_{desc}$) and increasing the dimensions of P^'($3 \times 2$) and S ($3 \times 3$).

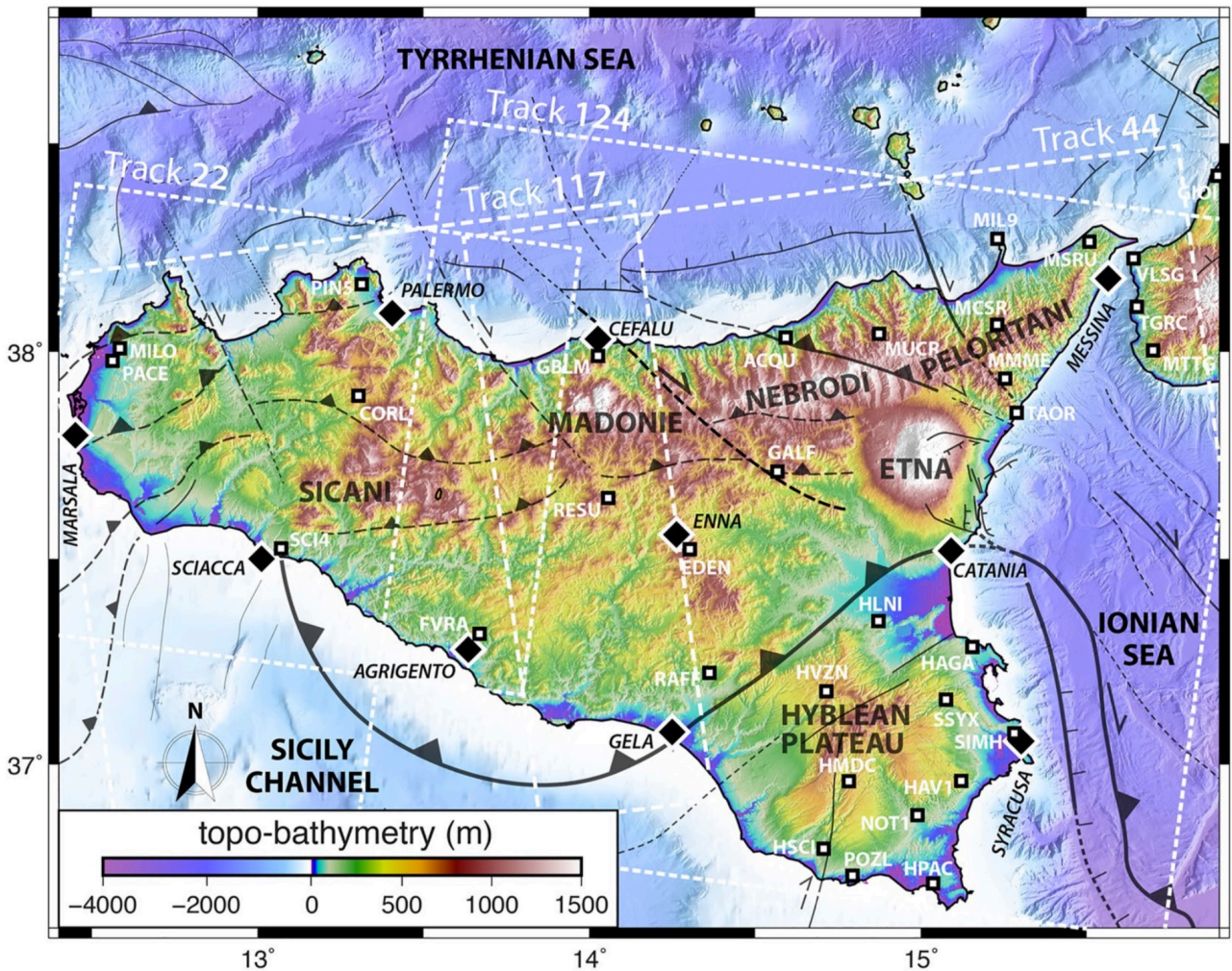

**Figure 2.** Shaded topo-bathymetric map showing the locations of the selected GNSS stations (white squares) and the Sentinel-1 tracks used in this study (Descending: 22 and 124/Ascending: 117 and 44). Topographic data are from the Japan Aerospace Exploration Agency (https://www.eorc.jaxa.jp) and the bathymetric compilation from Gutscher et al. (2016). Main Plio-Quaternary faults synthetized in Henriquet et al. (2020) are outlined in black.

## 2.2. PS-InSAR Processing

We processed all SAR images (TOPSAR Interferometric Wide swath acquisition mode) acquired by Sentinel-1 from the beginning of 2015 to the beginning of 2020 and distributed through the French platform PEPS. In order to cover the whole of Sicily, tracks 44 (sub-swaths 1 and 2) and 117 (all three sub-swaths) in ascending orbits as well as tracks 22 (sub-swaths 1 and 2) and 124 (all three sub-swaths) in descending orbits were selected (Figure 2). This leads to data sets of 196, 216, 205, and 215 radar images for tracks 44, 117, 22, and 124, respectively. The VV polarization was chosen since it has been determined empirically that it systematically provides higher SNR than the VH polarization.

The differential interferograms were generated using the Sentinel Application Platform (SNAP, Foumelis et al., 2018), which, in particular, allows for debursting images, pasting adjacent sub-swaths, merging consecutive acquisitions along an orbit, and subtracting the topographic phase contribution. Here, we use the SRTM digital elevation model at a 3 s sampling interval (~90 m). After careful checking, no significant topographic phase residuals were detected on individual differential interferograms. This is likely due to the relatively low topographic gradients in Sicily and the choice of the primary image selected so that all perpendicular baselines are <150 m (Figures S9 in Supporting Information S1). All interferograms belonging to a specific track share the same primary image, which has been chosen to minimize the temporal and geometric decorrelation. These primary images are 12/07/2017, 11/06/2017, 03/09/2017, and 01/05/2017 for tracks 44, 117, 22, and 124, respectively.

The PS processing of these interferograms is performed using the StaMPS software (Hooper et al., 2012). These analyses are mainly controlled by two parameters (see the Supporting Information for complete details). First, the amplitude dispersion index determines the initial selection of pixels as admissible PS candidates. This parameter is set to 0.3. Second, a statistical analysis of phase distribution is used to define a maximum percentage of selected pixels with a random phase. We set this later parameter to 5%. We chose these very strict parameter values to prioritize high-quality scatterers rather than spatial density. Nevertheless, such a choice yields to a dense spatial coverage of PS over most of the regions of Sicily. Recovering PS in poorly covered zones would require a large increase of these parameters, affecting the reliability of the velocity fields.

In low active tectonic settings, the atmospheric phase delay is one of the major biases disturbing interferometric phase interpretation (e.g., Beauducel et al., 2000). The PS-InSAR analysis can mitigate the influence of unwanted atmospheric phase delay patterns by specific spatiotemporal filtering based on its correlated signature in space but not in time. However, such an approach is much less efficient with relatively permanent atmospheric patterns. Moreover, removing modeled atmospheric phase delays can help with the sensitive unwrapping process. Hence, we incorporated atmospheric models for modeling their phase delays on all interferograms using the TRAIN software (Bekaert et al., 2015). Recent tropospheric models (e.g., ECMWF weather model) have considerably increased their spatial resolution compared to previous models (e.g., ERA-Interim database, Berrisford et al., 2011). In this study, we used the GACOS model (Yu et al., 2018), which combines the 0.125° and 6h resolution ECMWF weather model with tropospheric delays obtained from GNSS sites belonging to the study region. The use of GACOS significantly influenced the final mean velocity field, notably in the Peloritani-Calabria area where specific meteorological conditions (with some clear steadiness in time) occur. GACOS succeeded in precisely modeling them, while low spatial resolution models did not. The standard deviation of the phase in the interferograms is reduced by about 2–3 radians (Figure S10 in Supporting Information S1). We recognize that biases due to local atmospheric patterns may still exist at some locations, and one could try to evaluate them by computing the pseudo-3D velocities independently of each ascending and descending orbit. But testing these different geometries independently would require the use of additional data, such as the interpolated components of the GNSS velocity field, which may also induce local biases.

Steep slopes are found on Mt Etna as well as in the Peloritani and Madonie ranges. However, their influence in generating high topographic phase gradients is largely mitigated by the low values of perpendicular baselines. Consequently, no related phase aliasing was identified. Moreover, we incorporated atmospheric calibration prior to unwrapping, which leads to the reduction of phase gradients due to spatially changing tropospheric conditions. Therefore, the likeliest sources of unwrapping error are the significant eruptions on Mt Etna in December 2018 and May 2019. Although further exploration is required, a close look at the wrapped/unwrapped interferograms on Mt Etna did not reveal apparent unwrapping bias. If some biases remain, they are very likely spatially limited to the volcano itself.

## 2.3. PS Velocity Field Adjustments to 3D-GNSS Networks

In order to express the PS mean velocity fields along the LOS into a terrestrial reference frame, one needs a reliable 3D GNSS velocity field expressed in that reference frame. However, despite the good spatial coverage of GNSS sites all over Sicily (especially in the Eastern and most tectonically active parts of the island), no fully reliable 3D velocity fields have been published (Figure S1 in Supporting Information S1). Previous studies either provide only the horizontal component (e.g., Angelica et al., 2013; Mastrolembo Ventura et al., 2014) or the vertical component (Anzidei et al., 2021; Serpelloni et al., 2013) or they give 3D velocities but at a larger scale than Sicily (Masson, Mazzotti, Vernant, & Doerflinger, 2019). In this latter case, a close look at individual GNSS time series revealed that further processing was required to better take into account artifacts that very often alter the time series.

We retrieved GNSS position time series from the Nevada Geodetic Laboratory global solution (Blewitt et al., 2018) to generate an accurate and homogeneous 3D GNSS velocity field over Sicily. These positions are based on a global Precise Point Positioning solution using JPL products and aligned on the IGS14 reference frame (cf., http://geodesy.unr.edu/). We process all sites located in Sicily and in southwestern Calabria in order to extract reliable constant velocities by modeling and correcting annual and semiannual seasonal signals as well as instantaneous offsets (potential equipment change, earthquake, volcanic activity, or other origins, Figures S11 and S12 in Supporting Information S1). Time series modeling is performed using an inversion software developed at Geosciences Montpellier (Masson, Mazzotti, & Vernant, 2019). Several GNSS stations were discarded when their records are too short (e.g., AGRG in Agrigento) or affected by numerous and large temporal gaps (e.g., CPAN in Panarea Island). We also excluded stations located on Mt Etna (BELP, BRO2, and ESLN), since the corresponding time series are affected by significant transient displacements related to eruptive events.

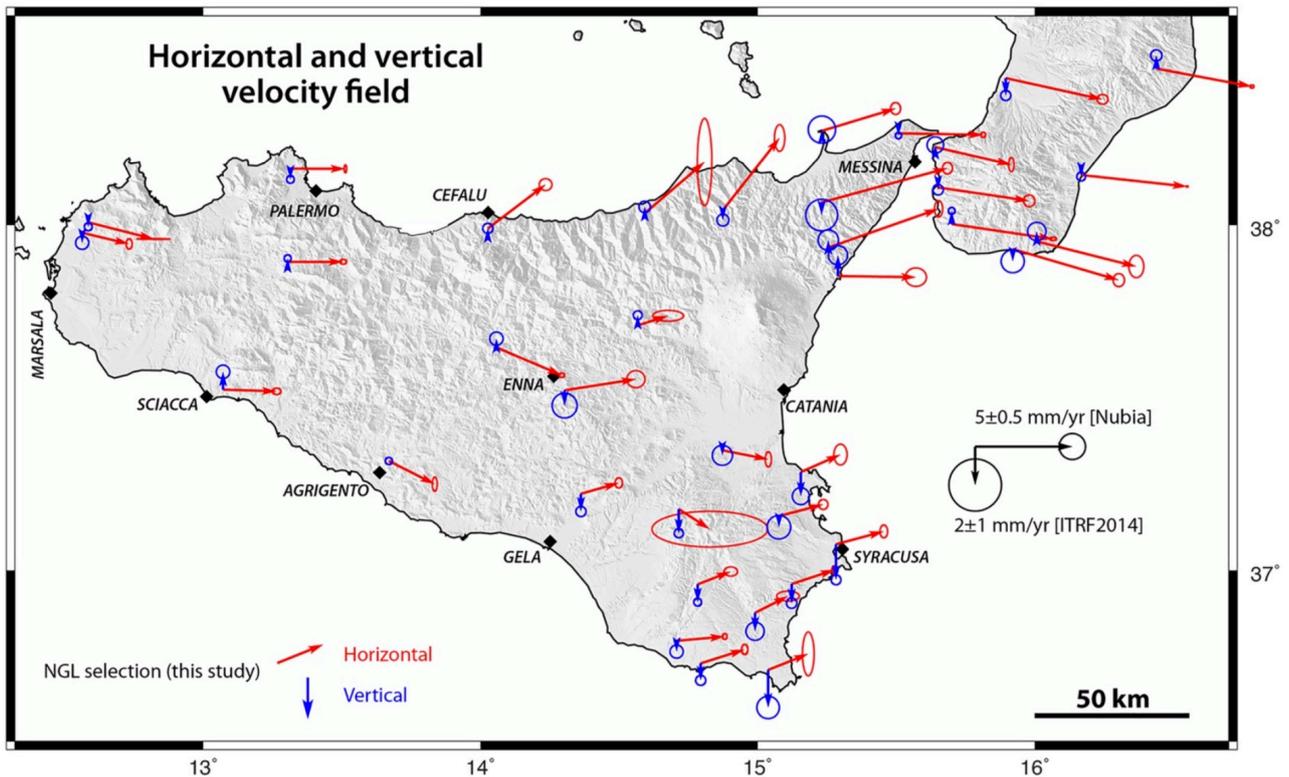

**Figure 3.** 3D velocity solution from our reanalysis of GNSS time series (red: Horizontal components relatively to fixed Nubia; blue: Up component in the ITRF2014 reference frame).

Theoretically, nothing prevents their incorporation in the adjustment process, as long as they span the same period as the PS measurements. Nevertheless, the estimation of the mean velocity may be more erratic (both for PS and GNSS). Considering that our objective is simply to constrain a velocity ramp at a large spatial scale and that we have tens of GNSS sites all over the rest of the island, we excluded them from our selection. We finally selected 40 sites (Figures 2 and 3) with reliable 3D velocities (Table 1 and Figure S1 in Supporting Information S1). This selection covers the whole Sicily Island with higher coverage along the eastern coast. The main difference with previously published GNSS velocity fields lies in the vertical component of some specific sites. This is notably the case for the GBLM site (Cefalù) that had a significant (and very likely unrealistic) upward value in Masson, Mazzotti, and Vernant (2019), Masson, Mazzotti, Vernant, and Doerflinger (2019), as well as the TGRC site (Reggio di Calabria), which had a significant downward value that has been confirmed neither by any of the two PS velocity fields, nor by our reprocessing of the position time series (Figure S5 in Supporting Information S1). In this study, we decided to express all velocity fields (PS and GNSS) in the Nubia horizontal reference frame and in the ITRF2014 vertical reference (Altamimi et al., 2016). PS mean velocity fields are primarily generated with a null mean value. To convert these velocity fields into the Nubia and ITRF horizontal and vertical reference frames, we correct the PS velocity fields from a spatial ramp that best fits (in a weighted least-square sense) the selected 3D-GNSS velocities projected along the local LOS with the PS velocities averaged in the vicinity of the GNSS sites. This adjustment becomes critical once we are concerned with the long-wavelength gradient of deformation as it is the case, for example, if one addresses the issue of ground deformation induced by the slab dynamics along northern Sicily. Hence, close attention must be paid to this adjustment on both PS and GNSS sides. However, while GNSS data may control part of the long-wavelength signals, the PS velocity field still provides independent short- and medium-wavelength measurements. For each GNSS site, a mean PS velocity value is estimated within the smallest radius containing at least 50 PS. Practically, this strategy provides enough PS for a reliable average estimation of their velocities, while staying close enough to the GNSS site (typically, a few hundred meters).

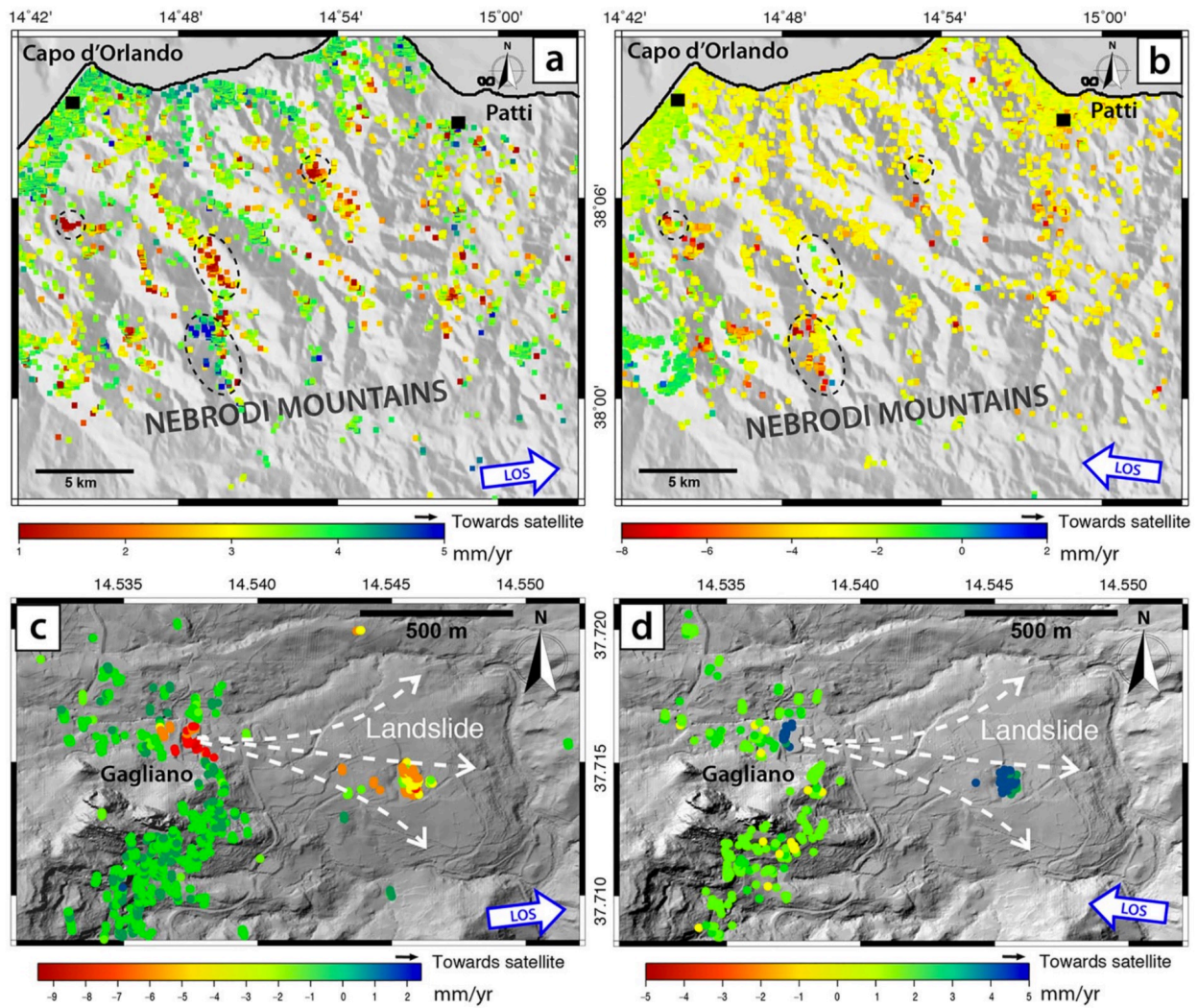

**Figure 4.** Mean Permanent/Persistent-Scatterer (PS) velocity fields are locally affected by transient processes, such as gravitational mass movements, localized from outliers in the ascending (a) and descending (b) line of sight velocity field over the Nebrodi range (dashed ellipses). PS mean velocities in ascending (c) and descending pass (d) close to GALF GNSS site (city of Gagliano Casteferrato) show a good agreement between PS velocity anomalies and the geomorphological signature of the Gagliano landslide.

The disparity of PS velocities (Figure S13 in Supporting Information S1) over this selection provides an estimate of the PS velocity uncertainty used in the weighted-fitting procedure. For each GNSS site, we checked that no bias is introduced by resampling a velocity gradient that would not be recorded by the GNSS site. This is notably the case in northeastern Sicily where numerous areas are affected by gravitational instabilities (Figure 4).

Some GNSS sites stand close to obvious ground instabilities. For instance, site GALF stands at ~1.5 km from the well-known Gagliano-Castelferrato landslide (e.g., Maugeri et al., 2006), while most of the selected PS for site MMME are located on the neighboring villages of Roccafiorita and Limina that are largely affected by gravitational mass movements. Most of these instabilities are identified from field observations. However, our PS velocity fields provide dense and precise cartography of these objects all over Sicily. Thereby, decreasing the search radius for sites affected by such processes allows for the selection of PS that we believe are now consistent with the GNSS measurement.

Still with the objective of comparing PS and GNSS measurements, we checked for the similarity of the time series of GNSS and selected PS. Three examples are presented in the Supporting Information S1 (Figures S11, S14, and S15 in Supporting Information S1). The PS time series are much less dense in time than the GNSS ones (6 days vs. daily solution) and the dispersion of raw PS positions (neither filtered in time nor in space) over time is high (a few tens of millimeters). Hence, it is generally difficult to draw definite conclusions about the existence of any faint transient processes over short time interval from PS time series, unless their amplitude is high as in the case of volcanic eruptions or anthropic deformations.

Considering that the selected GNSS sites are essentially steady, PS time series confirm, at first-order, (a) their stability (case of NOT1, site in Noto, Figure S14 in Supporting Information S1) and (b) that some discontinuities in GNSS time series are due to antenna changes rather than local ground deformations (case of TAOR, site in Taormina, Figure S11 in Supporting Information S1). In some cases, PS time series question the steadiness of the GNSS time series (case of GBLM, a site close to Cefalù, Figure S5 in Supporting Information S1).

The adjustment of PS velocity fields to the GNSS velocity measurements is performed separately for each satellite track. We checked afterward that the continuity of the mean velocity gradients is fulfilled, despite the difference in LOS at the spatial extremities of adjacent tracks that leads to different PS spatial locations and slightly different LOS velocities. We tested a joint adjustment of PS velocity fields to GNSS on both adjacent tracks with the constraint of minimizing the PS velocity difference on the overlapping area, but it did not change significantly the solution presented in this study.

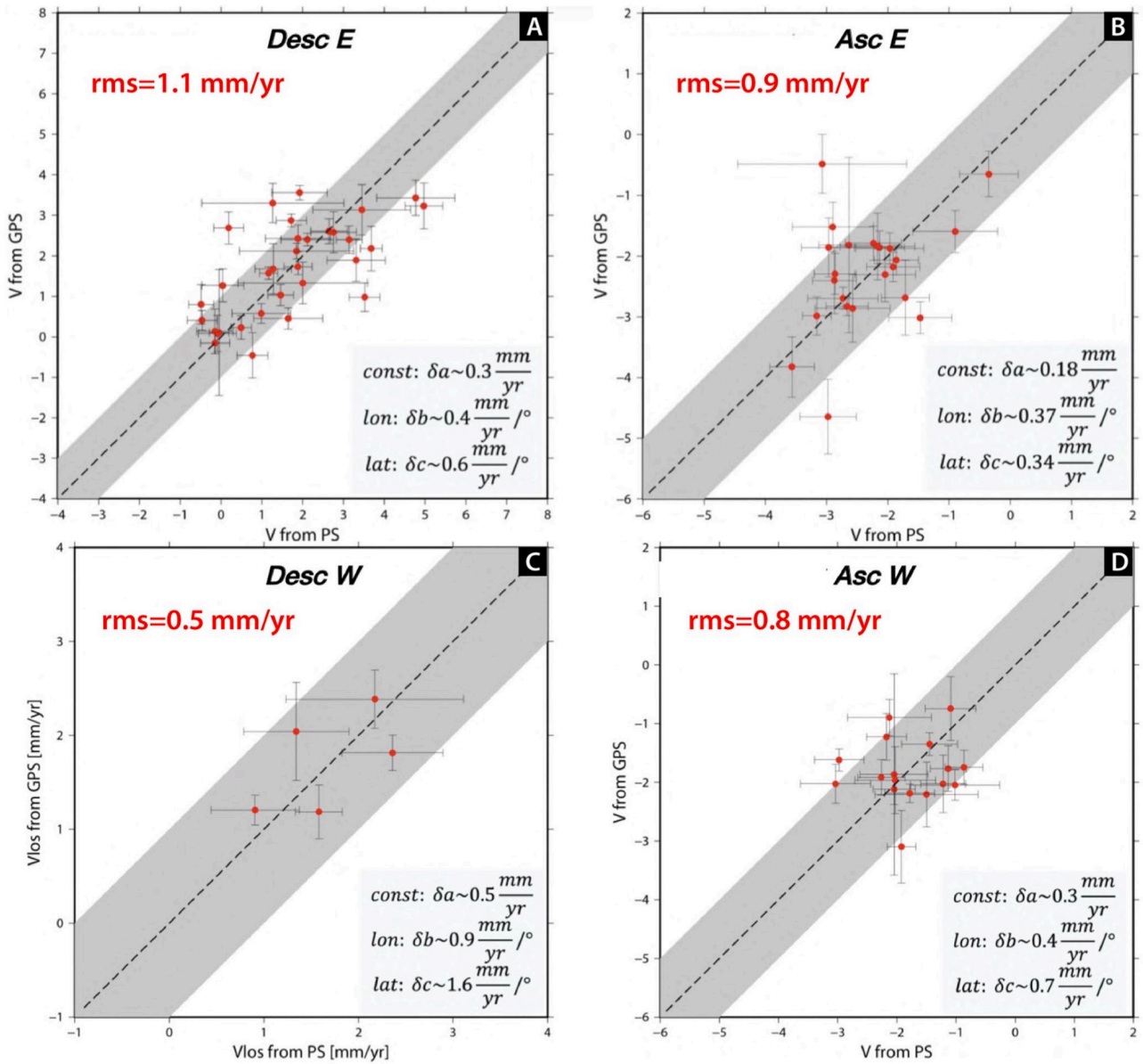

**Figure 5.** Comparison between GNSS velocities projected along the line of sight direction and Permanent/ Persistent-Scatterer mean velocities in their neighborhood: Eastern track (124) in descending (a) and (44) ascending (b) orbits, Western track (22) in descending (c) and (117) ascending (d) orbits. Ramp correction characteristics are indicated in the bottom-right corners (see also Figures S16, S17, S18 and S20 in Supporting Information S1). The 1-1 (±1) mm/yr relation is shown in gray, and RMS values for each plot are given in the upper-left corners.

The global agreement between the adjusted PS and GNSS LOS velocities is very good on all tracks: rms = 1.01, 0.48, 0.96, and 0.71 mm/yr, respectively, for descending tracks (east and west) and ascending tracks (east and west; Figure 5). The only sites that display a slight discrepancy (above 1 mm/yr) are TAOR, MUCR, and MCSR.

The last two stand within the Peloritani range, a vegetated area full of unstable ground surfaces (Figures 4a and 4b). In the case of MCSR, most of the selected PS in ascending orbit locate relatively far from the GNSS site (more than 1 km), while in the case of MUCR, it is unclear whether the PS are located in a stable area, considering the numerous ground instabilities affecting this region. Regarding TAOR, no explanation for this discrepancy can be found in the PS selection that would not be consistent with the GNSS since they are all in the center of the upper city of Taormina within about 100 m. We hesitated in using this GNSS site since its time series displays probable transient effects as well as discontinuity due to antenna changes (Figure S11 in Supporting Informa- tion S1). Moreover, it stops in mid-2018, which means that it covers neither the final time interval of PS, nor the main eruptions of Mount Etna in December 2018 and May 2019. Hence, the PS and GNSS mean velocity comparison for TAOR remains questionable. Regardless, all these discrepancies are small and do not alter the final ramps (Figures S16–S19 in Supporting Information S1) that are derived from these adjustments.

Finally, the uncertainties on the ramp parameters estimation rely on the spatial density and distribution of the GNSS sites as well as on the uncertainties associated with both the GNSS and mean PS velocities. The orders of magnitude for the ramp parameters uncertainties are about 0.2 mm/yr and 0.4 mm/yr/deg, respectively, for the constant and the longitudinal or latitudinal component. Only the western track in descending geometry has poorer uncertainties (0.5 mm/yr, 0.9 and 1.6 mm/yr/deg, respectively, for the constant, the longitudinal and the latitudinal components) since solely five GNSS sites lie in this area (Figure 5). Applying these ramp corrections (Figures S16–S19 in Supporting Information S1) provides the PS mean velocities that cover the whole Sicily Island both in ascending and descending geometries (Figure 6 and Figures S20–S23 in Supporting Information S1). Despite their strict selection, PS velocity fields in both passes cover the whole island without critical gaps. As expected, the PS density decreases in the highly vegetated mountainous region in northeast Sicily (Figure 6). Yet, the existence of numerous villages allows the broad detection of ground deformation throughout these areas. As attested by the PS/GNSS adjustment process, the agreement between the PS and GNSS along LOS velocities is very good. Even the MTTG site (south Calabria), which has been discarded from the adjustment procedure due to ground instabilities very close to the GNSS site, fits fairly well the regional trend of PS velocities. Nevertheless, moderate (but clear) discrepancies are found for stations MUCR, MSFR, MCSR, and MIL9 in the Peloritani range and TAOR just northeast of Mt Etna (Figure 6). As previously mentioned, the time series of all these GNSS sites can be questioned as well as potential bias of the PS velocity field within this densely vegetated region affected by numerous gravitational instabilities. Hence, definite conclusions in this area must be careful drawn. Longer time series both for PS and GNSS will help resolving such uncertainties.

## 2.4. Pseudo-3D Ground Velocity Field

Using Equation 1 and assuming that the North component of the velocity has been correctly removed from the PS LOS velocities, we derived a pseudo-3D mean velocity field (Figure 7) from the PS velocities in ascending and descending LOS directions (Figure 6). These results must be interpreted in light of the uncertainties (Figure S24 in Supporting Information S1) that arise from the diagonal elements of matrix $T$ (Equation 2). Uncertainties on the East and Up components of the reconstructed mean velocities are typically below 1 mm/yr, except in north-eastern Sicily where they reach 1–2 mm/yr, and at the top of Mount Etna where they can exceed 3 mm/yr, likely due to the frequent summit volcanic activity and winter snow cover. Note that Figure 7 represents mean velocities averaged over the 2015–2020 timespan, containing both long-term tectonic signals, such as inter-seismic loading or uplift, and transient displacements imposed by gravitational mass movements, anthropogenic subsidence, volcanic cycles, or hydrologic and climatic variations.

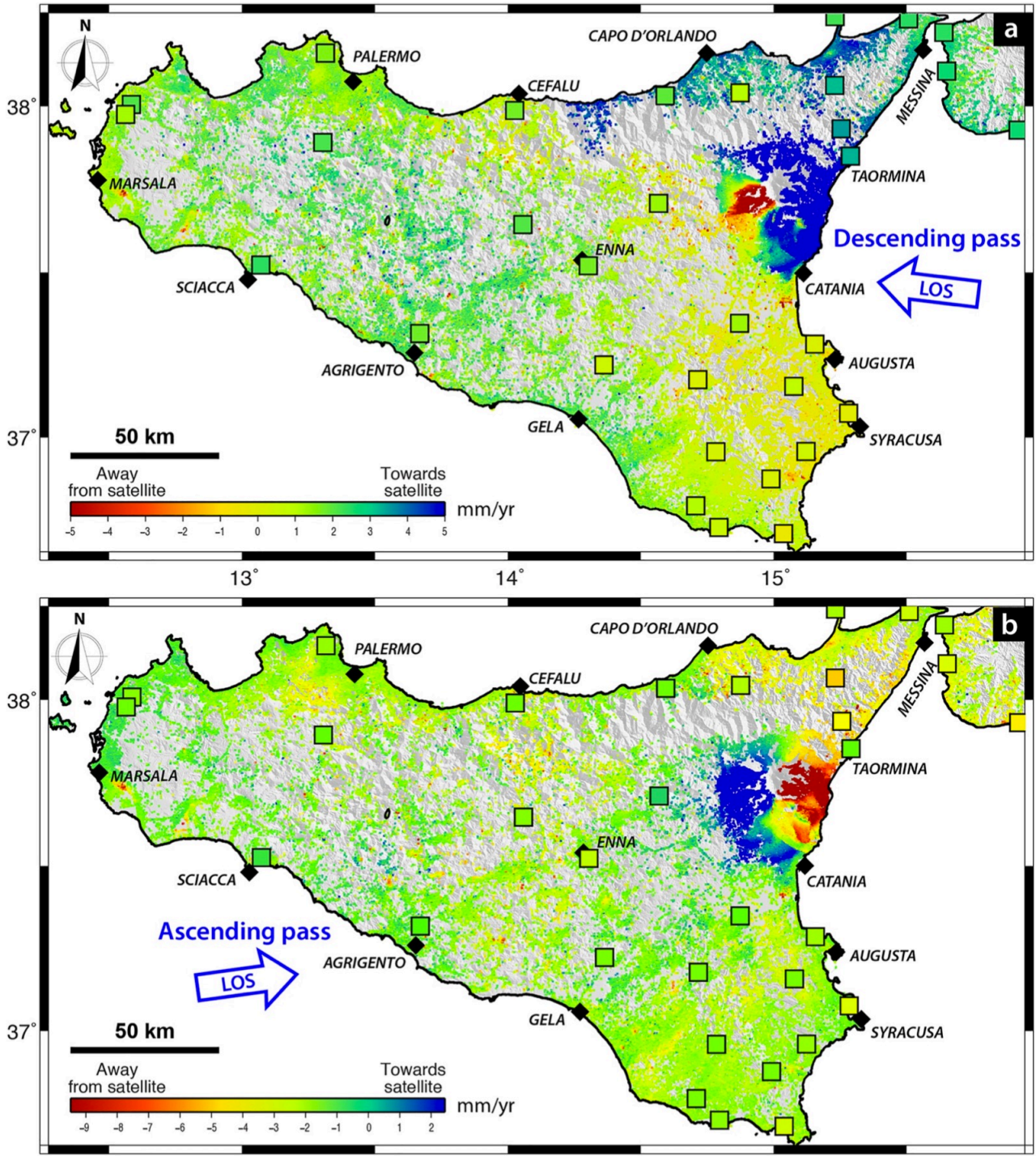

**Figure 6.** Comparison between the Permanent/Persistent-Scatterer mean velocities and the GNSS velocities projected along the line of sight direction. (a) Mixed tracks in ascending orbits. (b) Mixed tracks in descending orbits. Mean velocity uncertainties are given in the Figure S13 in Supporting Information S1.

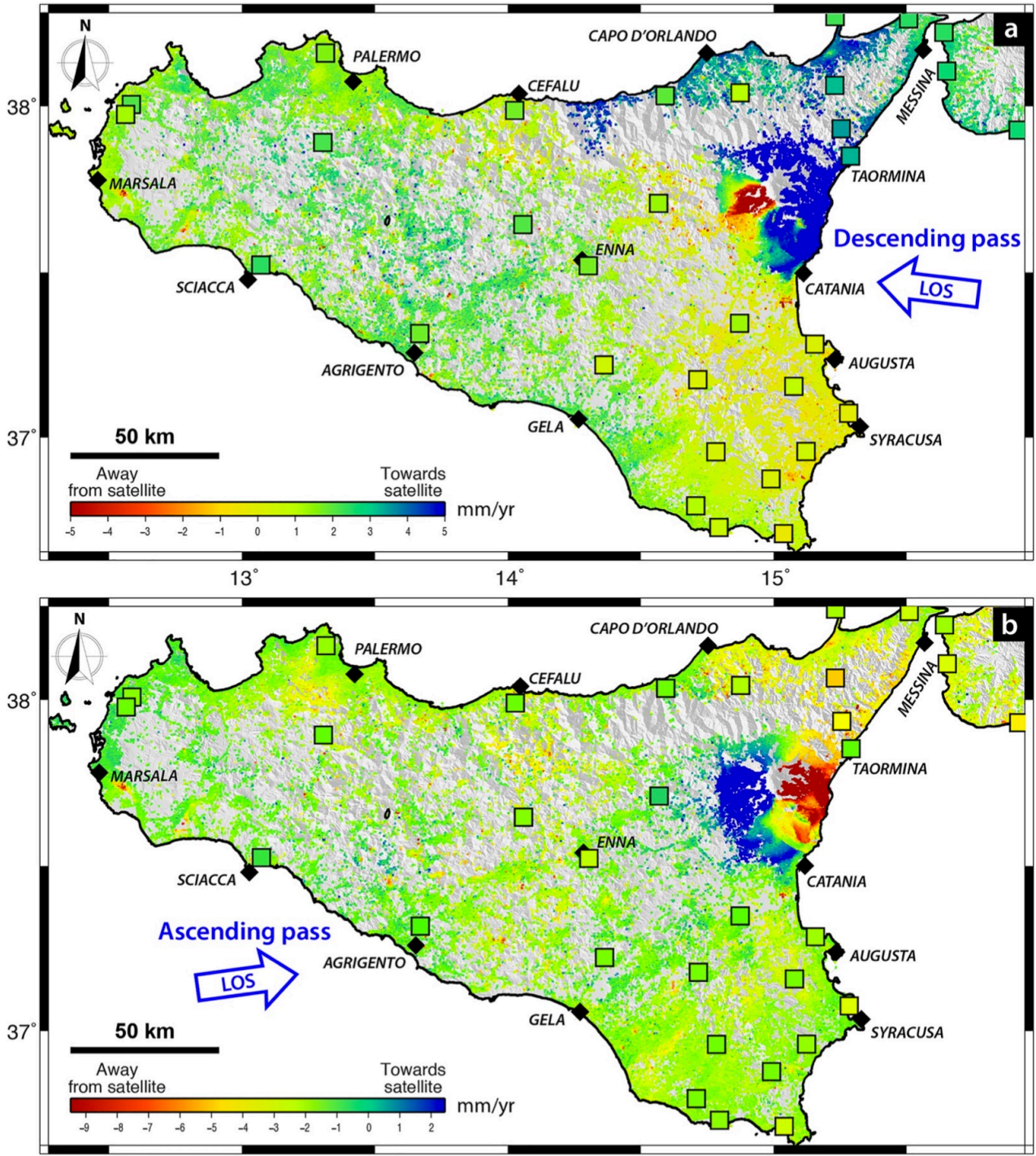

**Figure 7.** Vertical (a) and East-West (b) average surface velocity fields over the period 2015–2020 derived from Permanent/ Persistent-Scatterer data adjusted to GNSS mean velocities. Four zones of interest are detailed in the Figures 8–11. Mean velocity uncertainties are given in the Figure S24 in Supporting Information S1.

## 3. Interpretation of the East-Up Ground Velocity Field in Its Geological Context

Interpreting the crustal deformation and kinematics of Sicily is a difficult task because numerous geodynamic processes interact with each other over a few hundred kilometers, such as rifting in the Messina Strait and the Sicily Channel (e.g., Corti et al., 2006; D'Agostino & Selvaggi, 2004; Martinelli et al., 2019; Monaco, Tortorici, et al., 1996), back-arc spreading of the Tyrrhenian Sea (e.g., Faccenna et al., 1996, 2001; Rosenbaum & Lister, 2004), ongoing narrow subduction of the Ionian plate beneath Calabria (e.g., Maesano et al., 2017; Neri et al., 2009; Scarfì et al., 2018), potential slab tear along the Eastern Sicilian margin (e.g., Argnani, 2009; Dellong et al., 2018; Gallais et al., 2013; Gutscher et al., 2016; Maesano et al., 2020; Polonia et al., 2016; Wortel & Spakman, 2000), detached slab beneath northern Sicily (e.g., Argnani, 2009; Barreca et al., 2019, 2016; Wortel & Spakman, 2000), and the cessation of the Sicilian collisional system (Devoti et al., 2008).

The present study provides additional data to better characterize the current kinematics of Sicily. At first order, Figure 7 shows that most of the surface deformation is localized around Mount Etna and along the Cefalù-Etna seismic zone that separates the Nebrodi-Peloritani range from the Madonie Mounts and Central Sicily (Billi et al., 2010). A significant long-wavelength deformation signal is also identified across the Hyblean Plateau (southeastern Sicily), and several localized high subsidence patterns are clearly imaged in Western Sicily. Here-after, a closer look on these regions is provided (Figures 8–11) together with preliminary interpretations on the potential processes at the origin of the measured ground displacements.

### 3.1. Etna Region

The strong tectono-volcanic activity of Mount Etna causes significant ground deformations (Figure 8) that largely exceed the uncertainties of about 3–5 mm/yr near the top of the volcanic edifice (Figure S24 in Supporting Information S1). Such signal justifies why numerous studies have focused on the volcano dynamics using InSAR data (e.g., Bonforte et al., 2011; Borgia et al., 2000; De Novellis et al., 2019; Doin et al., 2011; Froger et al., 2001). Over the 2015–2020 timespan, the mean vertical velocities range from –5 mm/yr to over 8 mm/yr near the summit, and the East-component mean velocities range from –35 mm/yr to 35 mm/yr (Figure 8).

This new data set confirms most of the previously imaged and discussed surface displacements associated with the Etna volcano dynamics. For instance, the fault system accommodating the Eastern flank collapse has been active during the recorded 5 years. Along the Pernicana fault, up to 10 mm/yr of sinistral and 3 mm/yr of normal components are estimated in agreement with the local kinematics derived from field observations and geodetic measurements.

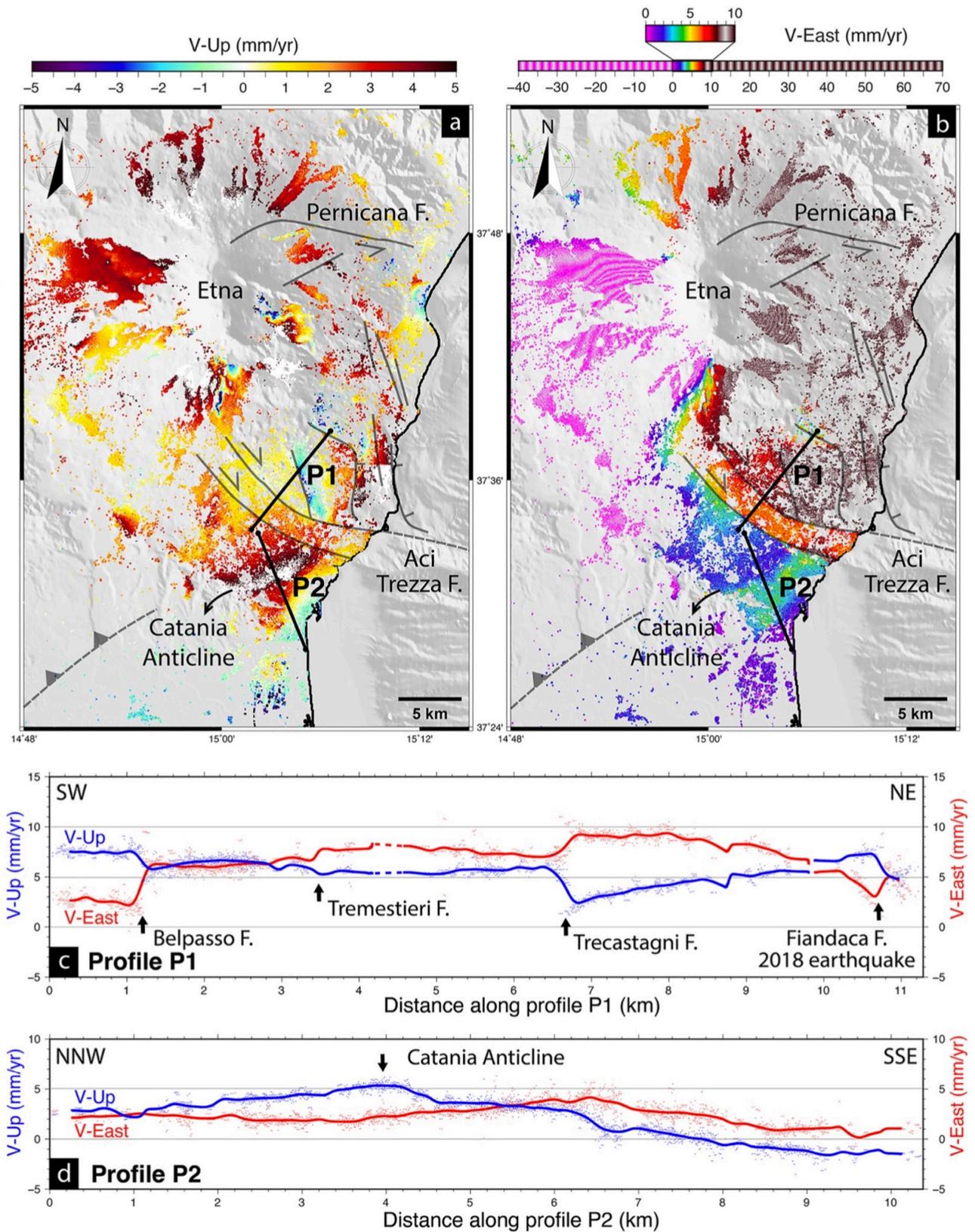

**Figure 8.** Vertical (a) and East-West (b) mean velocity fields derived from PS data at Mount Etna. A clear uplift of the edifice is visible on the Up component and the Etna eastward flank collapse is also clearly recognized on both components. Major known faults are outlined (black lines) as well as the Catania Anticline along the southern flank of Mount Etna. Vertical and East-West mean velocity profiles across the southern fault system (C, Profile P1) and the Catania Anticline (D, Profile P2) are extracted from the 3D mean velocity field. PS mean velocities are stacked within 200 m from each profile.

The creep rates derived from past studies are typically higher, 20–28 mm/yr and 10–20 mm/yr for the horizontal and the vertical components, respectively (Azzaro et al., 2001; Bonforte et al., 2007; D'Amato et al., 2017 and references therein). According to our measurements, the southern boundary of the sliding area also absorbed ~6 mm/yr of dextral slip distributed on the Belpasso, Tremestieri, and Trecastagni fault segments with locally up to 1–3 mm/yr of normal slip component (Figure 8c). These values are also lower than previous analysis that estimated ~15 mm/yr and ~5–6 mm/yr of dextral and normal slip rates from 1995 to 2000 (e.g., Bonforte et al., 2011).

It is interesting to note that the northern and southern fault systems controlling the eastern flank collapse are likely coupled since they show concomitantly lower creeping rate during the 2015–2020 timespan relative to previous studies. Moreover, deformation induced by the December 2018 earthquake is also recognized at the northeastern extremity of profile P1 (Figure 8c). Such kinematics illustrates the complex behavior of these tectono-magmatic fault systems with the alternation of seismic and aseismic deformation phases (e.g., Barreca et al., 2013; Mattia et al., 2015; Monaco et al., 2021).

From the presented data, the Catania Anticline, underlining the frontal thrusting of the Sicilian fold and thrust system (De Guidi et al., 2015; Ristuccia et al., 2013), has a clear kinematic expression with an ~4 mm/yr uplift relative to the linear trend of vertical rates along the NNW-SSE profile P2 (Figure 8d). This is consistent with the previous InSAR estimates covering the 1995–2000 period (Bonforte et al., 2011). This geodetic pattern coincides with long-term folding that started 240,000 years ago (Terreforti anticline, Ristuccia et al., 2013). Based on the deformation analysis of coastal-alluvial terraces, Ristuccia et al. (2013) estimated a mean uplift rate of 1.2 mm/yr for the last 240 ka, three times lower than the InSAR-derived uplift rates. It is not clear whether the current activity of the Catania Anticline is driven solely by the aseismic accommodation of the convergence along the front of the chain. One possible explanation for the InSAR/long-term discrepancy could be that the recent Etna volcano growth (mostly in the last 120,000 years, Barreca et al., 2018) interacts with the anticline activity through the inflation-deflation cycles of the edifice or as proposed by previous authors, induces the gravitational spreading of the southern flank of Mount Etna over the Plio-Quaternary clayey substratum (e.g., Bonforte et al., 2011; Borgia et al., 2000).

Finally, our data show a general uplift of Mount Etna at a rate of about 1–3 mm/yr that is consistent with the long-term continuous inflation of the edifice since 1990 as evidenced by local GNSS time series (e.g., Barreca, Branca, et al., 2020).

## 3.2. Northeastern Sicily

Significant PS velocity gradients are captured in Central-Eastern Sicily between the Madonie Mountains and the Peloritani Range (Figures 7 and 9). At first order, the reconstructed East and Up velocity fields show that the Nebrodi-Peloritani range is moving eastward at ~3 mm/yr and uplifting at ~1–2 mm/yr relative to the Central Sicily (Figure 9).

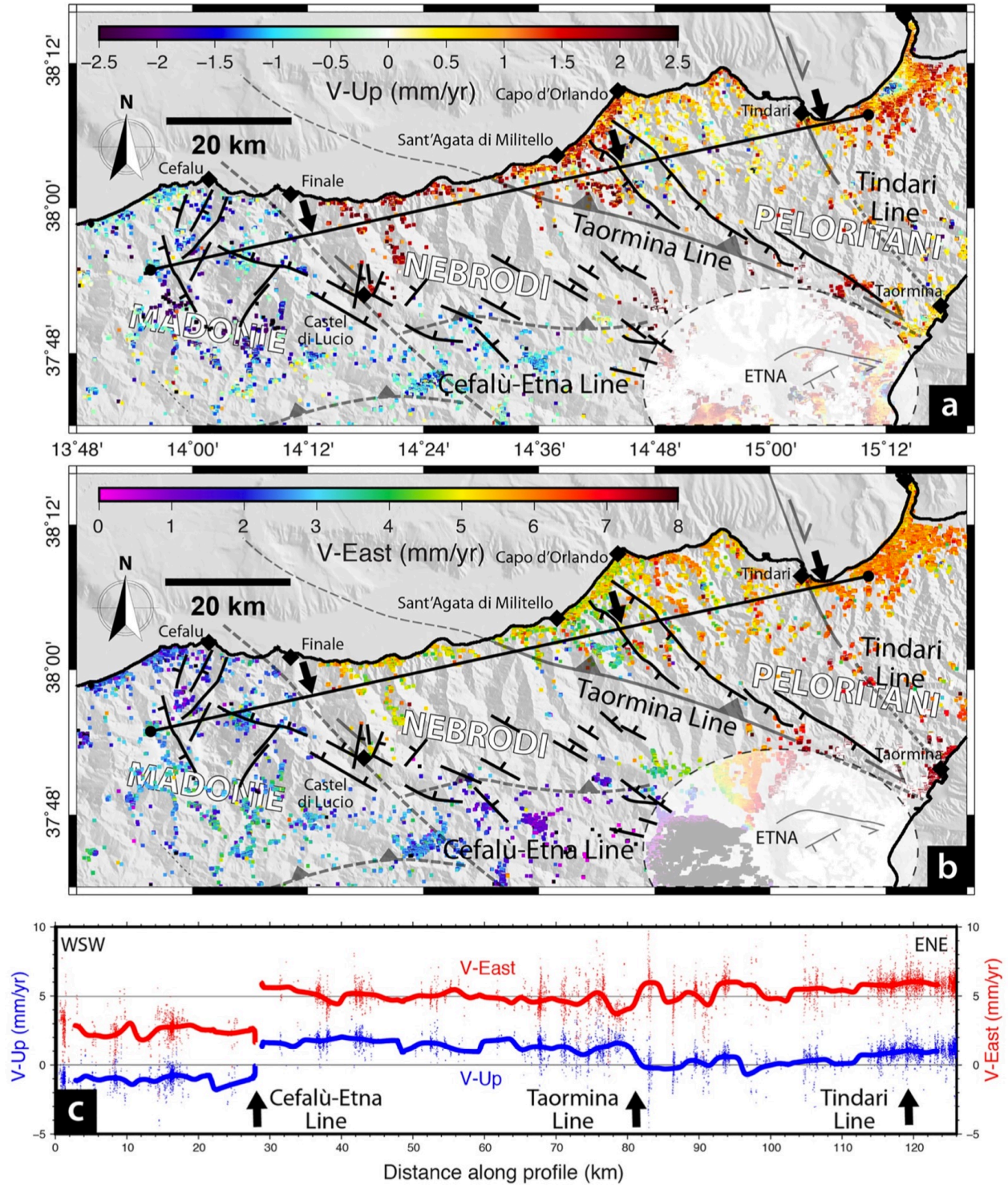

**Figure 9.** Vertical (a) and East-West (b) mean velocity fields from PS data along the northern coast from the Madonie Mountains to the Peloritani range. Main fault systems discussed in the literature are underlined in gray, such as the Tindari Fault System (e.g., Billi et al., 2006; De Guidi et al., 2013; Mattia et al., 2009; Scarfi et al., 2016), the Taormina Line (e.g., Pavano et al., 2015), and the Cefalù-Etna seismic zone (Barreca et al., 2016; Billi et al., 2010). Detailed black faults are from Billi et al. (2010) and Barreca et al. (2016) along the "Cefalù-Etna seismic zone" and from Pavano et al. (2015) along the Taormina Line. Vertical and East-West mean velocity profiles along the northern coast (c) are extracted from the 3D mean velocity field with PS mean velocities stacked within 2 km from the profile. Between the Madonie and Nebrodi range, the "Cefalù-Etna seismic zone" likely accommodates 2–3 mm/y of differential vertical displacements and 2–3 mm/yr along the East component.

The main velocity gradient is localized between the Madonie and Nebrodi mountains, where a velocity difference of 2–3 mm/yr in the East and Up components is identified roughly between Finale and Castel di Lucio within a <10-km-wide transition zone (Figure 9c). Such velocity gradients are greater than the mean velocity uncertainties in this area (about 1.5–2 mm/yr on the East component and 1–1.5 mm/yr on the Up component, Figure S24 in Supporting Information S1). This region accommodates upper-plate dextral shearing associated with the deep lithospheric tear induced by the retreat of the Ionian slab to the south-east. Previous field and geophysical studies (e.g., Barreca et al., 2016, 2019; Billi et al., 2010; Cultrera et al., 2017; Pavano et al., 2015) as well as GNSS data analysis (Bonforte & Guglielmino, 2008; Cultrera et al., 2017; Mattia et al., 2009; Palano et al., 2012; Scarfì et al., 2016, 2018) have identified three active tectonic belts: (a) the Tindari Fault System (Billi et al., 2006), slicing from the Aeolian Islands (Lipari-Volcano) to Tindari and Capo S. Alessio further south; (b) the Taormina Line between San'Agata di Militello and Taormina; and (c) the "Cefalù-Etna Seismic zone" (Billi et al., 2010), running toward the south-east of Cefalù (Figures 9a and 9b). The Tindari Fault System extends from the Lipari-Vulcano complex (central Aeolian Arc) to the Ionian offshore and possibly acts as a STEP fault related to the Ionian rollback (e.g., Barreca et al., 2016; Barreca, Branca, et al., 2020; Gutscher et al., 2016; Maesano et al., 2020). The Taormina Line delim- its the contact between the Peloritan block and the accretionary wedge units of the Nebrodi Mounts (Henriquet et al., 2020 and references therein). The "Cefalù Seismic zone" follows, in its northern part, a potential Pliocene STEP fault that extended to the southeast up to the Judica Mountains (e.g., Barreca et al., 2016). According to our PS-InSAR analysis, the "Cefalù-Etna Seismic Zone" appears as the major active zone in this region, although most studies point to the Tindari Line and sometimes to the Taormina Line. Our analysis is consistent with the intense seismic activity (Figure 1b) along the "Cefalù-Etna seismic zone" (Billi et al., 2010, Figures 9a and 9b), but also with the GNSS-based block models from Mastrolembo Ventura et al. (2014). While the discontinuity in the PS velocity field does not correlate with a known significant seismogenic structure affecting the surface, it is broadly consistent with SE-trending extensional active fault segments mapped in the area (Billi et al., 2010, Figures 9a and 9b).

The Peloritani range is delimited to the East by a NE-SW-oriented extensional domain, the Messina Strait (Figures 1 and 7), where the 1908 Messina earthquake occurred. The Strait of Messina has developed in the Calabro-Peloritani upper plate since the Late Pliocene-Early Pleistocene (Monaco, Tortorici, et al., 1996) and can be explained by multiple mechanisms (Doglioni et al., 2012), such as: (a) back-arc extension, (b) faster retreat of Calabria together with the southeastward rollback of the Ionian slab relative to the eastern Sicilian margin, and (c) lithospheric doming in response to slab tear or break-off (see also Barreca et al., 2021). Our reconstructed East and Up velocity fields do not show any significant vertical differential velocities in this region. However, an E-W-distributed velocity gradient of 2–3 mm/yr is quantified between the southwestern part of the Peloritani range and Western Calabria. The comparison with past GNSS studies, suggesting 1.5–3.4 mm/yr of extension in the WNW-ESE to NW-SE direction (e.g., D'Agostino & Selvaggi, 2004;

Serpelloni et al., 2010), is made difficult by the lack of the north component of the InSAR horizontal velocity field.

## 3.3. Hyblean Plateau

The East and Up velocity fields in Southeastern Sicily (Figures 7 and 10) benefit from low mean PS velocity uncertainties (0.5–1 mm/yr, Figure S24 in Supporting Information S1), but their interpretations are challenged by the lack of localized velocity gradients. At a large scale, the eastern part of the Hyblean Plateau is subsiding relative to the western part at a rate of ~2 mm/yr (Figures 10a and 10c). On the ENE-trending velocity profiles, a small short-wavelength velocity gradient is visible in the East component (Figures 10b and 10c) at the inter-section with the Scicli-Ragusa and Comiso-Chiaramonte fault systems, but the velocity difference (~1 mm/yr) is close to the PS uncertainties in this area (~1 mm/yr, Figure S24 in Supporting Information S1). Nevertheless, D-InSAR analysis covering the 2003–2010 period has also indicated a possible reactivation of these fault systems (Vollrath et al., 2017).

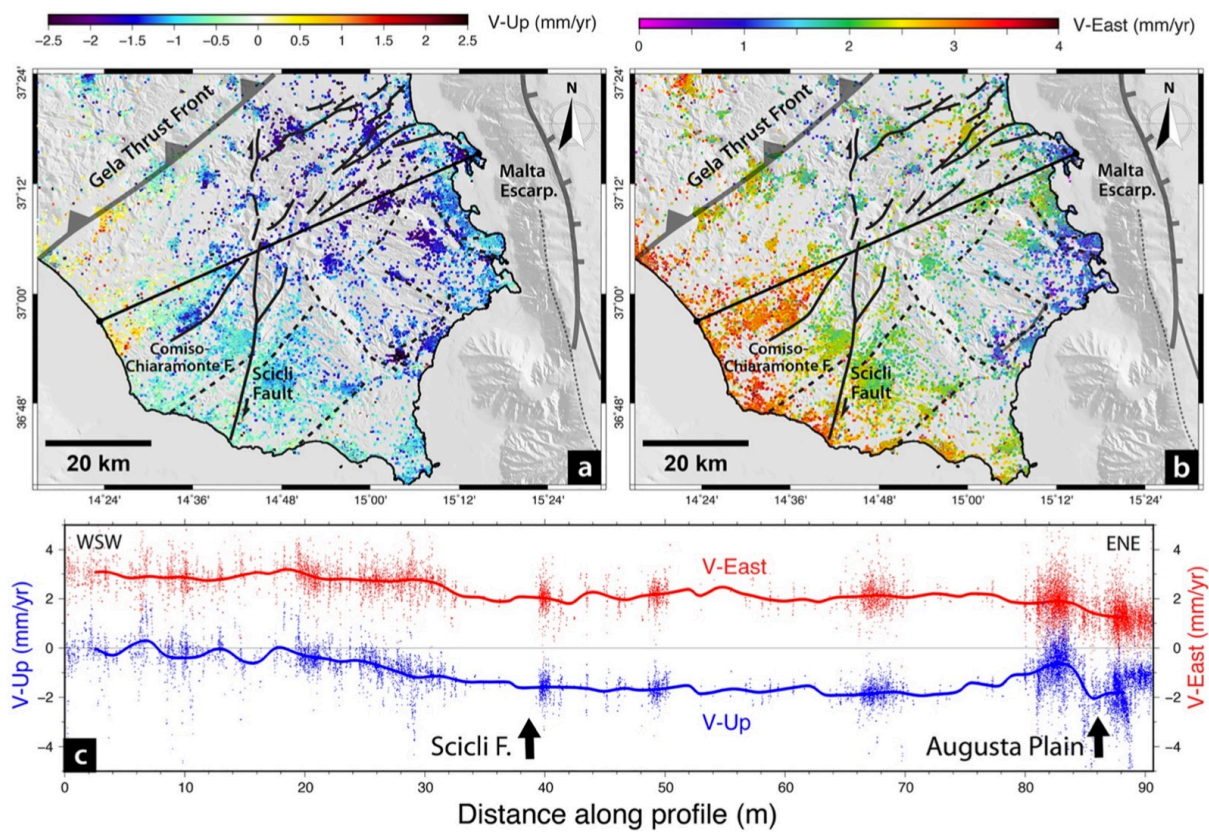

**Figure 10.** Vertical (a) and East-West (b) mean velocity fields from Permanent/Persistent-Scatterer (PS) data in the Hyblean Plateau. Main fault systems discussed in the literature are shown in black (simplified from Cultrera et al., 2015) as well as the inferred Gela Thrust Front, separating the Hyblean foreland from the Sicilian orogenic wedge, and the Malta Escarpment bordering the eastern coast (thick gray). Vertical and East-West mean velocity profiles across the Hyblean Plateau (WSW-trending black line from Borgo Alcerito to Augusta) are extracted from the 3D mean velocity field with PS mean velocities stacked within 2 km from the profile. A clear subsidence of the eastern Hyblean Plateau of ~−2 mm/yr relative to the western part is observed.

Along the eastern border of the Hyblean-Malta platform, the NNW-trending Malta Escarpment separates the Pelagian continental realm from the Ionian oceanic and stretched continental domain (Oceanic-Continental Transition) along an ~ 250-km-long and >2500-m-high bathymetric scarp. This major morpho-structural feature is inherited from the Mesozoic rifting (e.g., Frizon de Lamotte et al., 2011), but is likely reactivated along its northern portion north of Siracusa (Gambino et al., 2021) in response to the Ionian lower plate down-flexing (Argnani & Bonazzi, 2005) or the slab tear (e.g., Argnani, 2009; Govers & Wortel, 2005). We speculate that the relative 1–2 mm/yr of subsidence of the eastern Hyblean Plateau may have a tectonic origin in contrast to the very localized strong subsidence patterns related to water pumping detectable in the area (Anzidei et al., 2021; Canova et al., 2012). The large scale (~100 km in wavelength) subsidence pattern could be consistent with the inter-seismic elastic deformation signal expected on the footwall of the Malta Escarpment, a major fault system that likely corresponds to the seismogenic source of the Noto 1693 earthquake (e.g., Argnani & Bonazzi, 2005; Gambino et al., 2021).

In the northern and western parts, the Hyblean Plateau is partly underthrusted below the Gela Thrust Front (GTF). Along this major structural contact, no surface deformation can be identified in the East and Up mean velocity fields (Figure 10), even though published GNSS data evidence a N-S-oriented contraction pattern along the northern border of the Hyblean Plateau (4–5 mm/yr; Mattia et al., 2012; Palano et al., 2012). Such kinematics is consistent with the geological data, suggesting that the western border of the GTF is sealed by Middle-Late Pleistocene sediments (Lickorish et al., 1999), whereas along the northern border, contraction is accommodated by thrusting and folding (e.g., De Guidi et al., 2015; Ristuccia et al., 2013).

### 3.4. Central and Western Sicily

Central and Western Sicily appear relatively stable, particularly in the East component of the PS velocity field where no significant velocity gradients are detected (Figure 7b). The Western and Central sectors of Western Sicily are moving eastward relative to Nubia at the same rate of 3 ± 1 mm/yr, which is consistent with previous geodetic measurements (Serpelloni et al., 2010). Given the low uncertainties on the PS mean velocities in the Up component (<1 mm/yr, Figure S24 in Supporting Information S1), the entire region shows no significant vertical motion in the ITRF 2014 vertical reference frame, except in the Agrigento-Ribera area where a slight uplift of about 0.5–1 mm/yr is estimated (Figure 7a). At a local scale, the PS data reveal high subsidence rates, above −8 mm/yr, in particular around the Strasatti and Castelvetrano areas (Figure 11a).

Central and Western Sicily regions have a relatively low seismicity rate (Figure 1b), yet they have hosted destructive earthquakes as evidenced by the M~6 1968 Belice earthquake sequence (e.g., Bottari, 1973) and the destruction of the Selinunte Greek temples between 370 BCE and 600 CE (Bottari et al., 2009). The low instrumental seismicity of this region illuminates a regional south-verging thrust system (Lavecchia et al., 2007) compatible with focal solutions available for the 1968 earthquake (Anderson & Jackson, 1987). Structural field data point out an active NNW-dipping blind thrust as the source of the Belice compressive seismic events (Monaco, Mazzoli, & Tortorici, 1996). However, the presented East and Up mean velocity fields do not evidence significant deformation in the epicentral area of the Belice earthquake (Figure 11). Such lack of observable ground deformation in this known tectonically active zone is not surprising, given the short InSAR temporal coverage and low strain rates in the area.

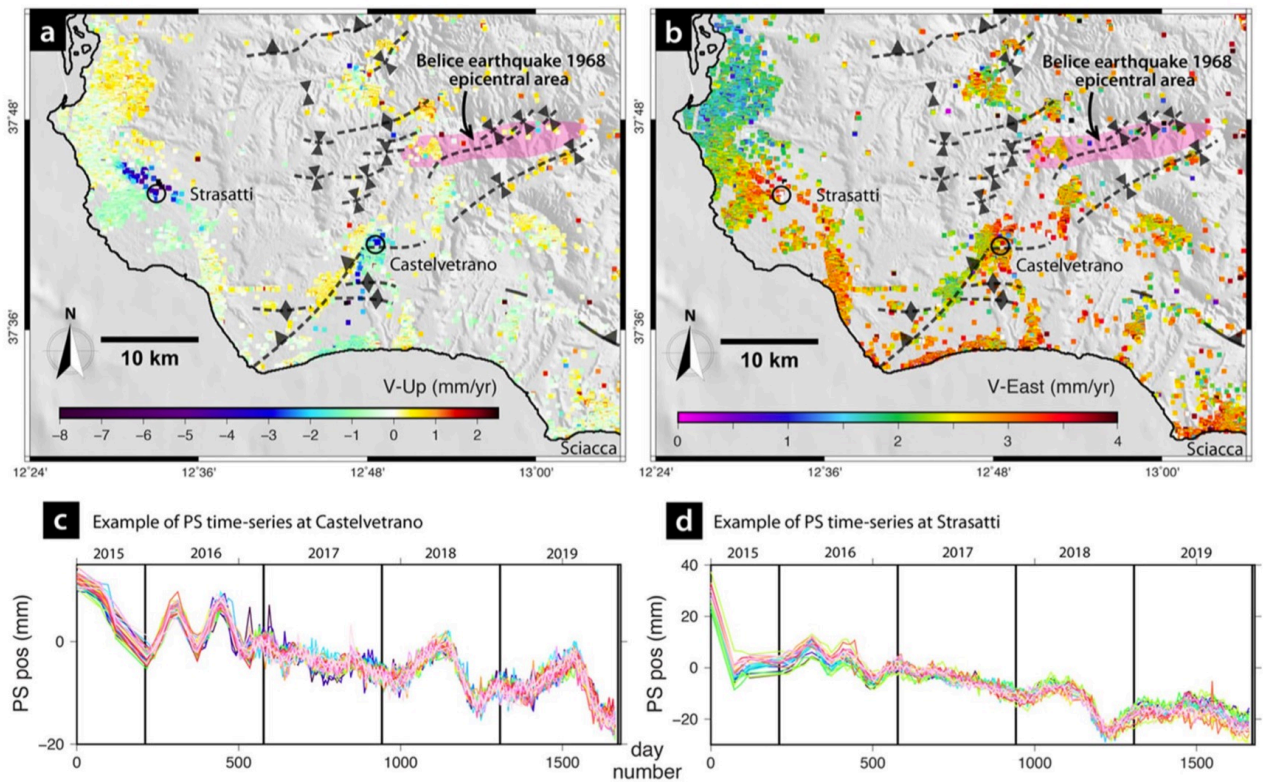

**Figure 11.** Vertical (a) and East-West (b) mean velocity fields from Permanent/Persistent-Scatterer (PS) data in southwestern Sicily. Main fault systems and folds shown in black are from Barreca et al. (2014). The gray ellipsoid corresponds to the most damaged area (total damage >80%) during the 1968 Belice earthquake (from Monaco, Mazzoli, & Tortorici, 1996; Monaco, Tortorici, et al., 1996). Two significant patterns of subsidence are identified at Strasatti and Castelvetrano. Time series of PS positions in descending line of sight direction are extracted (within the black circles) in both subsiding area around Strasatti (c) and Castelvetrano (d). Note the similar trends and variations in both places, possibly due to the combination of tectonic, anthropic, and hydrologic-climatic changes.

To the southwest of the Belice region, the fast subsidence at Castelvetrano and Strasatti were also clearly identified in previous works using Differential-InSAR analysis based on ENVISAT images acquired between 2003 and 2010 (Barreca et al., 2014; Barreca, Bruno, et al., 2020). These authors interpreted the Strasatti subsiding area as related to groundwater pumping based on reported aquifer monitoring (Barreca et al., 2014 and references therein), and they proposed a tectonic origin for the Castelvetrano-Campobello di Mazara differential velocity due to damaged archeological relics in the area. Although we agree that there is a clear velocity gradient along the Castelvetrano-Campobello di Mazara alignment coinciding with the NE-SW trending active "CCA" fault (Barreca et al., 2014; Ferranti et al., 2021), more dedicated work is needed to extract the geodetic signature of the CCA fault. The horizontal deformation rate across the CCA fault is very low (<1 mm/yr, Barreca, Branca, et al., 2020; Barreca, Bruno, et al., 2020), indicating that this region undergoes a slow tectonic activity as confirmed by its low level of instrumental seismicity and the very long return period of reported historical earthquakes (De Lucia et al., 2020). On the contrary, the vertical velocity is very fast compared to the horizontal, up to 5 mm/yr and locally higher, which is likely out of range for the creeping fault hypothesis. In the Campobello di Mazara plain, groundwater overexploitation is likely as important as in the Strasatti area with an annual average recharge lower than 15% and an average decrease in the piezometric level of 20 m between 1981 and 1999 (Aureli et al., 2007 and references therein). The hypothesis of hydrological changes favoring subsidence appears also supported by the very similar trends of the PS time series in the Strasatti and Castelvetrano area (Figures 11c and 11d). We hypothesize that the CCA fault could act as a permeability barrier, driving the aquifer water flux and control part of the subsidence in the Castelvetrano-Campobello di Mazara plain.

## 4. Discussion

One critical issue for PS-InSAR analysis in low-deforming areas is to express the PS mean velocities in ascending and descending orbits into regional East-West and vertical velocity fields, here expressed in the Nubia and ITRF 2014 horizontal and vertical reference frames. The very good agreement between GNSS and PS mean velocities along the LOS (<1 mm/yr, Figure 5) on all tracks provides robust orbital ramp corrections, which ensures reliable East and Up reconstructions, assuming that the North component has been properly removed from the PS velocity fields. This assumption is likely locally wrong at Mount Etna, but does not invalidate the main conclusions on its dynamics.

We also note that in addition to the mean velocity uncertainties in the Up component (0.5–1.5 mm/yr, except at the top of Mount Etna), the ITRF2014 vertical reference frame has an accuracy of about 1 mm/yr when considering vertical velocities (Altamimi et al., 2016). Consequently, the low vertical displacement rates (–0.5 to 0.5 mm/yr) must be critically interpreted in terms of absolute subsidence or uplift patterns.

Our analysis benefits from the recent GACOS tropospheric model at high spatial (0.125°) and temporal (6h) resolution (Yu et al., 2018). Steady local meteorological conditions are well modeled by GACOS, leading to a reduction of the phase distribution in some interferograms of about 2–3 radians (Figure S10 in Supporting Information S1). Nevertheless, we cannot rule out remaining local atmospheric biases in the coastal mountainous areas like the Madonie, Nebrodi, Peloritani, Calabria, or Etna mountains. Recent studies have also shown that the Mediterranean region is prone to multiyear groundwater fluctuations in karst environments related to climatic cycles as the Northern Atlantic Oscillation that could have an impact on the estimated vertical velocities about ±1 mm/yr (Silverii et al., 2016). The processed Sentinel-1 time series are currently too short to model these transient deformations, but future works will likely improve the PS mean velocity fields by modeling seasonal and multiyear hydro-climatic signals.

The provided East-West and vertical velocity fields correspond to averaged time series from 2015 to 2020. Consequently, we pay special attention on the meaning of these velocities in terms of steady-state or transient processes. As shown by previous InSAR studies in Nebrodi-Peloritani range (Ciampalini et al., 2016; Saroli et al., 2005), gravitational mass movements are common processes that are also found in Central and Western Sicily within the sedimentary accretionary wedge units. It can be noted that the steep slopes of the Nebrodi-Peloritani water-sheds and the global uplift of this region are also consistent with the hillslope destabilization that affects the area. Fortunately, these effects are clearly identified and have a limited spatial extension (Figure 4). Main groundwater and mining areas are also relatively well identified in the vertical velocity field (Figure 7) as the subsidence rates significantly exceed the regional mean vertical velocity field. We discussed the case of Strasatti and Castelvetrano (Figure 11), but other areas are concerned in Sicily (Barcellona Pozzo di Gotto, Catania and Augusta plain, and Noto).

The deformation events associated with the magmatic activity of Etna (Figure 8) fluctuate with the eruptive cycles and cannot be interpreted as long-term continuous signals. Geodetic measurements on Mount Etna are extremely abundant (GNSS, InSAR, strainmeters, and tiltmeters), and their analyses have been extensively used to investigate the seismovolcanic activity and for documenting the lateral collapse of the edifice (Bonforte et al., 2011; Borgia et al., 2000; De Novellis et al., 2019; Froger et al., 2001; Mattia et al., 2015 among many others). Notably, continuous tilt data provide unique insights on rapid magma intrusions (Aloisi et al., 2020) rather than constraining long-term deformations that are difficult to quantify due to instrumental drifts (Furst et al., 2019). Such additional constraints could be used to better estimate the adjustment of our PS velocity fields or to compare both measurements. However, because the deformation of Mount Etna is spatially very heterogeneous due to these transient effects, its analysis deserves a specific study that is not attempted here.

Tectonic signals imaged by the East and Up PS velocity fields are difficult to characterize due to the low-deformation setting of Sicily. Only two areas could be associated with active tectonic processes with reasonable confidence. We showed that the Nebrodi-Peloritani range diverges from mainland Sicily at a rate of 2–3 mm/yr along the East component with clear shear strain localization along the Cefalù-Etna seismic zone. These measurements, together with field evidence of fault activities (Figure 9, Barreca et al., 2016; Billi et al., 2010) associated with a seismogenic lineament (Figure 1b), are all in favor of an active right-lateral transtensional transfer zone.

In order to better address the issue of tectonic deformation in such a vegetated area, the application of the SBAS approach should fill the gap of measurements between the PS presented in this study. Nevertheless, the existence of villages ensures the detection of PS from place to place throughout the NE mountain range. Despite their low spatial density (Figures S25–S28 in Supporting Information S1), we believe they are numerous enough to capture the large-scale tectonic deformation that affects this area. Extrapolating short-term satellite measurements to the long term is another important but challenging issue (Lamb, 2021). To do so, we compared in the Figure 12 the PS vertical velocity field with coastal Late Pleistocene (MIS 5.5 highstand, ~125 ka) and Holocene averaged vertical rates based on relative sea level changes (from Ferranti et al., 2006, 2010).

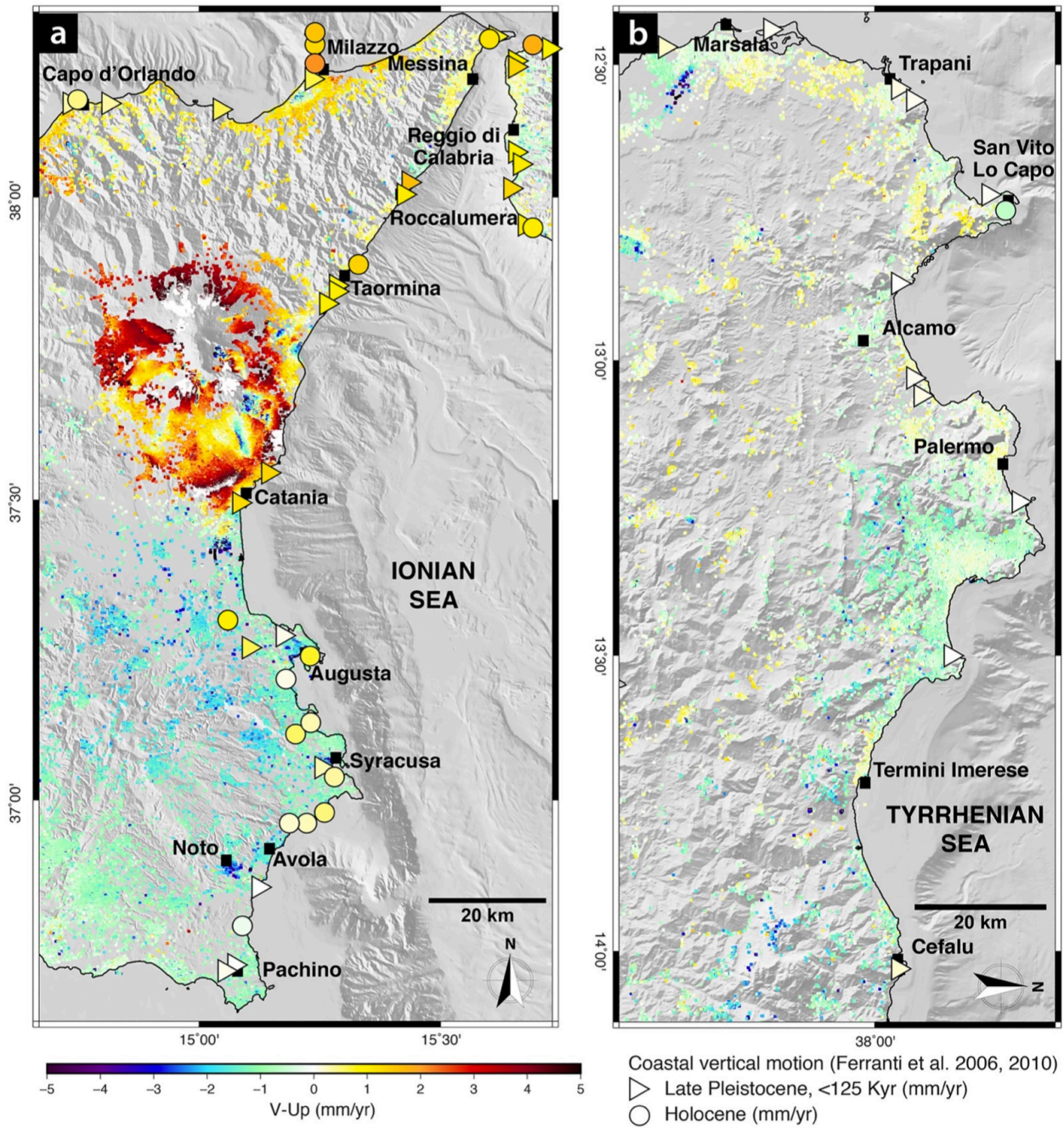

**Figure 12.** Vertical mean velocity fields from Permanent/Persistent-Scatterer (PS) data compared to Holocene (triangles) and Late Pleistocene (inverted triangles) coastal vertical rates along the eastern (a) and northwestern (b) coast (from Ferranti et al., 2006, 2010). Note the relatively good agreement between the coastline estimates and the PS vertical mean velocities in northwestern Sicily and northeastern Sicily, while the eastern Hyblean Plateau is clearly diverging from the long-term gentle uplift recorded by coastal data.

These long-term velocities are in agreement with the low activity of Western Sicily (0 ± 0.5 mm/yr, Figure 12b). In northeastern Sicily, Late Pleistocene and Holocene mean velocities show a significant uplift of 1 ± 0.5 mm/yr also consistent with the PS-InSAR data (Figure 12a), except along the Messina strait where we do not image any significant uplift. Such consistency between short-term and long-term vertical rates in northern Sicily suggests that long-wavelength ductile processes play an important role in the present-day kinematics.

In the eastern Hyblean region, the discrepancy between the geodetic subsidence and the long-term moderate uplift (from Ferranti et al., 2006, 2010) is of the order of 2–3 mm/yr over a 50-km-long transect along the coast (Figure 12a). As discussed in the previous section, we speculate that this short-term subsidence pattern, also evidenced by Anzidei et al. (2021), could represent a transient deformation related to the inter-seismic elastic loading of an east-dipping normal fault system located offshore the Hyblean Plateau, likely along the reactivated Malta Escarpment (e.g., Argnani & Bonazzi, 2005; Gambino et al., 2021). However, InSAR monitoring of inter-seismic loading phases on active faults is mostly limited to major strike-slip faults, such as the North Anatolian, the San Andreas, or the Haiyuan faults (e.g., Cavalié et al., 2008; Tong et al., 2013; Wright et al., 2001), where slip rates exceed several tens of mm/yr. In our case, we face the problem that the hanging wall is buried below the Ionian Sea, and the deformation rates of the eastern margin faults are very low (from 0.1 mm/yr to a maxi-mum of 3–7 mm/yr, Gambino et al., 2021), which question the ability of GNSS and PS to measure such type of deformation.

## 5. Conclusion

We have produced the first Sicily island-wide pseudo-3D mean velocity field from Sentinel-1 high spatiotemporal resolution acquisitions from 2015 to 2020. In order to convert the PS velocities in LOS direction into regional Nubia and ITRF2014 horizontal and vertical reference frames, we used GNSS time series carefully reprocessed from the regional GNSS networks. After incorporating GACOS tropospheric corrections, we obtain a very good agreement between the PS and the GNSS mean velocity fields, a necessary condition to interpret subtle deformations in this low-strain region. Our analysis provides, first, a mean velocity field along the East and Up components with possible tectonic velocity gradients reflecting long-term deformation rates, but also transient geological, hydrological, or anthropic processes that cannot be extrapolated over long time periods. From the good match between the PS velocity gradients along the Cefalù-Etna belt, the seismic activity, and the associated field evidences of active faults in the area, we confirm that the Nebrodi-Peloritani range is decoupled from main-land Sicily in response to the Ionian slab retreat. Our analysis does not show any significant strain localization along the Taormina Line and the Tindari Fault Systems to the East at least during the timespan covered by the Sentinel-1 data. In regard to the later, geodetic, seismological, and morpho-structural data revealed that it has been active in recent times, but the short PS time series are not appropriate for analyzing low inter-seismic strain on non-creeping faults. We also speculate from the disagreement between our geodetic analysis and the Late Pleistocene to Holocene vertical velocity rates that the currently subsiding Eastern Hyblean margin could sign the interseismic elastic loading of the Malta Escarpment, capable of generating magnitude >7 earthquakes, such as the Noto 1693 event. We therefore encourage future geodetic analyses in low-strain areas to integrate long-term constraints when interpreting potential tectonic signals. Finally, our analysis open doors to further interpretations of Sentinel-1 InSAR data for understanding transient phenomena, in particular tectono-magmatic processes, gravitational mass movements, hydrological variations, or anthropic local ground displacements.

## Data Availability Statement

All Sentinel-1 SAR images are available for download from the French platform PEPS (accessible at https://peps.cnes.fr/rocket/#/home). The time series analysis software StaMPS (accessible at https://github.com/dbekaert/StaMPS) comes from Hooper et al. (2012). The atmospheric corrections were performed using the GACOS model (accessible at http://www.gacos.net/) and incorporated using the TRAIN software from Bekaert et al. (2015) (accessible at https://github.com/dbekaert/TRAIN). All data sets and materials underlying this study are published open access in Henriquet et al. (2021; https://doi.org/10.5281/zenodo.5769120) and additional details are provided in Supporting Information.

## Acknowledgments:

We acknowledge the European Space Agency (ESA) and the Centre National d'Etudes Spatiales (CNES) that provided the Sentinel-1 radar images freely accessible from the French platform PEPS. Financial support comes from the CNRS-INSU TelluS-Aleas program. The maps and graphics presented in this study were generated using the Generic Mapping Tools (GMT) software (Wessel & Smith, 1998). GNSS time series were analyzed using the R software (https:// www.r-project.org). We are grateful to Mario Mattia and Christine Masson for the useful discussions about the GNSS measurements. We wish to thank Simon Lamb and two other anonymous reviewers for their constructive comments that significantly improved the manuscript

## Author Contributions:

**Data curation:** Maxime Henriquet, Michel Peyret, Stéphane Mazzotti
**Formal analysis:** Maxime Henriquet, Michel Peyret, Stéphane Dominguez
**Funding acquisition:** Stéphane Dominguez
**Investigation:** Maxime Henriquet, Michel Peyret, Stéphane Dominguez
**Methodology:** Maxime Henriquet, Michel Peyret, Stéphane Dominguez
**Resources:** Stéphane Dominguez
**Software:** Michel Peyret **Visualization:** Maxime Henriquet
**Writing – original draft:** Maxime Henriquet, Michel Peyret
**Writing – review & editing:** Stéphane Dominguez, Giovanni Barreca, Carmelo Monaco